\documentclass[fleqn,10pt]{wlscirep}
\usepackage[utf8]{inputenc}
\usepackage[T1]{fontenc}
\usepackage{xcolor}
\usepackage{url}
\usepackage{graphicx}
\usepackage{subcaption}
\usepackage{multirow}
\usepackage{multicol}
\usepackage{colortbl}
\usepackage[export]{adjustbox}
\usepackage{array}
\title{Towards End-to-End Structure Solutions from Information-Compromised Diffraction Data via Generative Deep Learning}

\author[1,*]{Gabe Guo}
\author[1]{Judah Goldfeder}
\author[2]{Ling Lan}
\author[1]{Aniv Ray}
\author[1]{Albert Hanming Yang}
\author[3]{Boyuan Chen}
\author[2,*]{Simon J. L. Billinge}
\author[4,*]{Hod Lipson}
\affil[1]{Columbia University, Department of Computer Science, New York, NY, USA 10027}
\affil[2]{Columbia University, Department of Applied Physics and Applied Mathematics, New York, NY, USA 10027}
\affil[3]{Duke University, Department of Mechanical Engineering and Materials Science, Durham, NC, USA 27708}
\affil[4]{Columbia University, Department of Mechanical Engineering, New York, NY, USA 10027}
\affil[*]{Corresponding authors: gzg2104@columbia.edu, sb2896@columbia.edu, hl2891@columbia.edu}

\begin{abstract}
The revolution in materials in the past century was built on a knowledge of the atomic arrangements and the structure-property relationship.  The \textit{sine qua non} for obtaining quantitative structural information is single crystal crystallography.
However, increasingly we need to solve structures in cases where the information content in our input signal is significantly degraded, for example, due to orientational averaging of grains, finite size effects due to nanostructure, and mixed signals due to sample heterogeneity.  
Understanding the structure property relationships in such situations is, if anything, more important and insightful, yet we do not have robust approaches for accomplishing it.
In principle, machine learning (ML) and deep learning (DL) are promising approaches since they augment information in the degraded input signal with prior knowledge learned from large databases of already known structures.  
Here we present a novel ML approach, a variational query-based multi-branch deep neural network that has the promise to be a robust but general tool to address this problem end-to-end.  
We demonstrate the approach on computed powder x-ray diffraction (PXRD), along with partial chemical composition information, as input. We choose as a structural representation a modified electron density we call the Cartesian mapped electron density (CMED), that straightforwardly allows our ML model to learn material structures across different chemistries, symmetries and crystal systems. 
When evaluated on theoretically simulated data for the cubic and trigonal crystal systems, the system achieves up to $93.4\%$ average similarity with the ground truth on unseen materials, both with known and partially-known chemical composition information, showing great promise for successful structure solution even from degraded and incomplete input data.
The approach doesn't presuppose a crystalline structure and the approach are readily extended to other situations such as nanomaterials and textured samples, paving the way to reconstruction of yet unresolved nanostructures.
\end{abstract}
\begin{document}

\flushbottom
\maketitle

\section*{Introduction}

% Why you should care
Crystallography is the experimental science of determining the structure of crystals by analyzing x-ray, neutron or electron diffraction patterns\cite{crystallography_giacovazzo2002fundamentals, crystallography_hammond2015basics, lipson1935crystal}. Powder crystallography is a sub-branch of crystallography that solves this problem when the measured sample consists of a large number of small, randomly oriented grains of the material \cite{simon_powder_crystallography_dinnebier2008powder, henry_lipson_daniel1943x, henry_lipson1942structure, lipson1984study}.  
This problem is mathematically harder because of the loss of orientational information which must be recovered through inference during the structure reconstruction. 
It is useful when single crystals are difficult to obtain experimentally.  
However, it also is a good starting point for developing methods to determine the structure of nanomaterials and molecules in solution \cite{billinge2007problem}, problems that currently have no robust solution. 

%\sjb{this is true but not relevant to the current problem...refinemnet of known structures is used for this work.....}The information provided by powder crystallography can illuminate much more than just structure: It can provide clues as to defects and phase transitions, and can be used to identify and compare materials, or to search for new compositions, for example 

% Problem statement
%The \textbf{powder crystallography problem, however, is much harder to solve} because interference patterns are averaged over every possible orientation of the crystal \cite{simon_powder_crystallography_dinnebier2008powder}. 
% , \textcolor{red}{exacerbating the well-known phase problem} \cite{billinge2007problem, crystallography_giacovazzo2002fundamentals, crystallography_hammond2015basics}.
%An exact analytical solution for powder crystallographic inference (\textit{i.e.}, determining crystal structures from XRD patterns) is not known to exist (although there is an analytical solution mapping crystal structures to XRD patterns). 
The field of structure determination from powder diffraction \cite{davidStructureDeterminationPowder2008a} has grown by adapting conventional crystallographic methods to the powder case.  As with all crystallographic methods, these use inference and an iterative design approach to obtain structure candidates.
%This process is standard even for observation modalities other than XRD, such as pair distribution functions (PDF) \cite{yangStructureminingScreeningStructure2020i, banerjeeClusterminingApproachDetermining2020i, lindahlchristiansenStructureAnalysisSupported2020}.
The approach is a human-intensive activity requiring hands-on guidance by skilled experts. It involves first identifying the crystallographic coordinate system, a process called indexing, followed by finding the fractional coordinates of atoms in the unit cell from Bragg peak intensities \cite{crystallography_giacovazzo2002fundamentals, davidStructureDeterminationPowder2008a}.
For PXRD data, the process sometimes works and sometimes does not, depending on the quality of the data and the complexity of the structure.
It is not a straightforward process and requires considerable expertise.

% Information that leads us towards our solution
Recent work suggests that deep learning methods hold great potential to simplify the solution of complex inference problems with a straightforward end-to-end process. 
For instance, the protein-folding problem has recently been "solved" by end-to-end deep learning approaches like AlphaFold \cite{alphafold_jumper2021highly, alphafold2_bryant2022improved} and RoseTTAFold \cite{baek2021accurate}. 
This is highly relevant, because protein folding is a sister problem to powder crystallography -- both problems involve recovering the enigmatic shape of complex molecules from sparse and low-dimensional (i.e., 1-dimensional) inputs (amino acid sequences for the case of proteins and PXRD patterns for the powder crystallography case) %\sjb{I changed RNA to DNA here because I htink that is correct, but I am on shaky ground, so splease check. I find it unlikely that it would be RNA which is intermediate step in protein replication, not the long-lived code.  But I may be wrong. }
\cite{dobson2003protein_fold}. Other examples of problems that have yielded to end-to-end learning are image classification \cite{alex_krizhevsky2012imagenet}, autonomous vehicle driving \cite{self_driving_bojarski2016end}, and speech recognition \cite{amodei2016deep_speech}.

% Why previous solutions fall short
Machine and deep learning methods have been proposed to accelerate various stages of the powder crystallographic process. However, most of these works are conducted in a classification or feature regression paradigm: given an observation such as the XRD pattern, predict a property of the structure, such as space group symmetry, phase, unit cell parameters, or magnetism \cite{liu;aca19, xrd_class_oviedo2019fast, xrd_class_ml_suzuki2020symmetry, xrd_park2017classification, xrd_lee2020deep, xrd_aguiar2019decoding, class_ziletti2018insightful, xrd_tiong2020identification, garcia2019learning, merker2022machine}. 
There are some works that generate crystal structures, but their methodologies are not readily applicable to our problem because they (1) largely focus on unconditional (with respect to XRD pattern) generation cases in which there is no ground truth structure to reconstruct \cite{deepmind_generative_merchant2023scaling, materials_diffusion_yang2023scalable, gflownet_hernandez2023crystal}; (2) solve the easier single-crystal diffraction problem \cite{pan2023deep, dun2023crysformer}; (3) were designed only for specific classes of materials, such as proteins \cite{alphafold_crystallography_barbarin2022x} and monometallic nanoparticles \cite{kjaerDeepStrucStructureSolution2023a}. Furthermore, the source code for many works in the deep learning for crystallography paradigm is not open-sourced, limiting their reproducibility \cite{pan2023deep, dun2023crysformer, materials_diffusion_yang2023scalable, class_ziletti2018insightful}.

Here, we propose an approach towards an end-to-end deep neural network that is able to determine a transformed  three-dimensional electron density field directly from a 1-dimensional diffraction pattern.
The actual electron density distribution may then be recovered with the inverse transform as we describe below.

% Our solution
The model we call \textit{CrystalNet} is a variational \cite{kingma2013auto_vae} query-based multi-branch deep neural network (DNN) architecture (also known as a conditional implicit neural representation \cite{yu2021pixelnerf, mildenhall2021nerf, tancik2020fourier, sitzmann2019scene, park2019deepsdf}) that takes powder x-ray diffraction patterns and chemical composition information as input, and outputs a continuous function that is related to the 3D electron density distribution. 
We call this function the Cartesian mapped electron density (CMED) because we map the electron density from the crystallographic coordinate system of the structure to a Cartesian coordinate system. 
This distorts the resulting electron density but places it on a universal basis that allows the model to be seamlessly trained on structures from different crystal systems and with different unit cell parameters.
The advantage of this representation for material structure is that it frees us from traditionally predefined properties such as the number of atoms and the crystallographic coordinate system. 
The actual electron density distribution may be recovered from the CMED through the inverse mapping, and if required,
the discrete molecular structure can be straightforwardly decoded from this electron distribution if needed \cite{hoffmannDataDrivenApproachEncoding2019}. 
After training, given a new, previously unseen diffraction pattern (and corresponding chemical composition information), \textit{CrystalNet} can be queried  to produce a 3D CMED map at any desired resolution. 
Due to our variational approach \cite{kingma2013auto_vae, higgins2016beta_vae}, \textit{CrystalNet} can also be queried multiple times to produce different predictions, should the first guess be sub-satisfactory. 
The design, training and testing protocols are described in the Methods section.

% Our results
The performance of the model are described here. We report preliminary results from the cubic and trigonal crystal systems using theoretically simulated data from the Materials Project \cite{materials_project_jain2013commentary}. \textit{CrystalNet} was able to reconstruct atomic structures from the cubic system almost perfectly. 
For the trigonal system, \textit{CrystalNet} achieves success in most cases, with the infrequent failure modes providing insights for future work. 
We experiment on these two crystal systems because their intra-crystal axial lengths are equal, thereby eliminating the need to predict lattice vector magnitudes. Although other crystal systems were not explored fully in this study, the results on these two crystal systems indicate that our approach can be highly effective for the remaining five systems.  
We note that the model does not make use of any  symmetry or chemical property information and yet still shows success.   
This means that such information may be added as priors in future iterations when there is even greater information loss in the input signal, for example, due to very low symmetry structures or broad diffraction signals charateristic of nanomaterials.

We also conduct ablation studies by systematically reducing input chemical composition information to gain insight into which information is most important for AI-enabled powder crystallography going forward. 
We find that while this information helps our model, for these high symmetry structures, crystal reconstruction is generally successful with only the XRD data and no compositional information at all.

\begin{figure}
    \centering
    \includegraphics[width=0.9\linewidth]{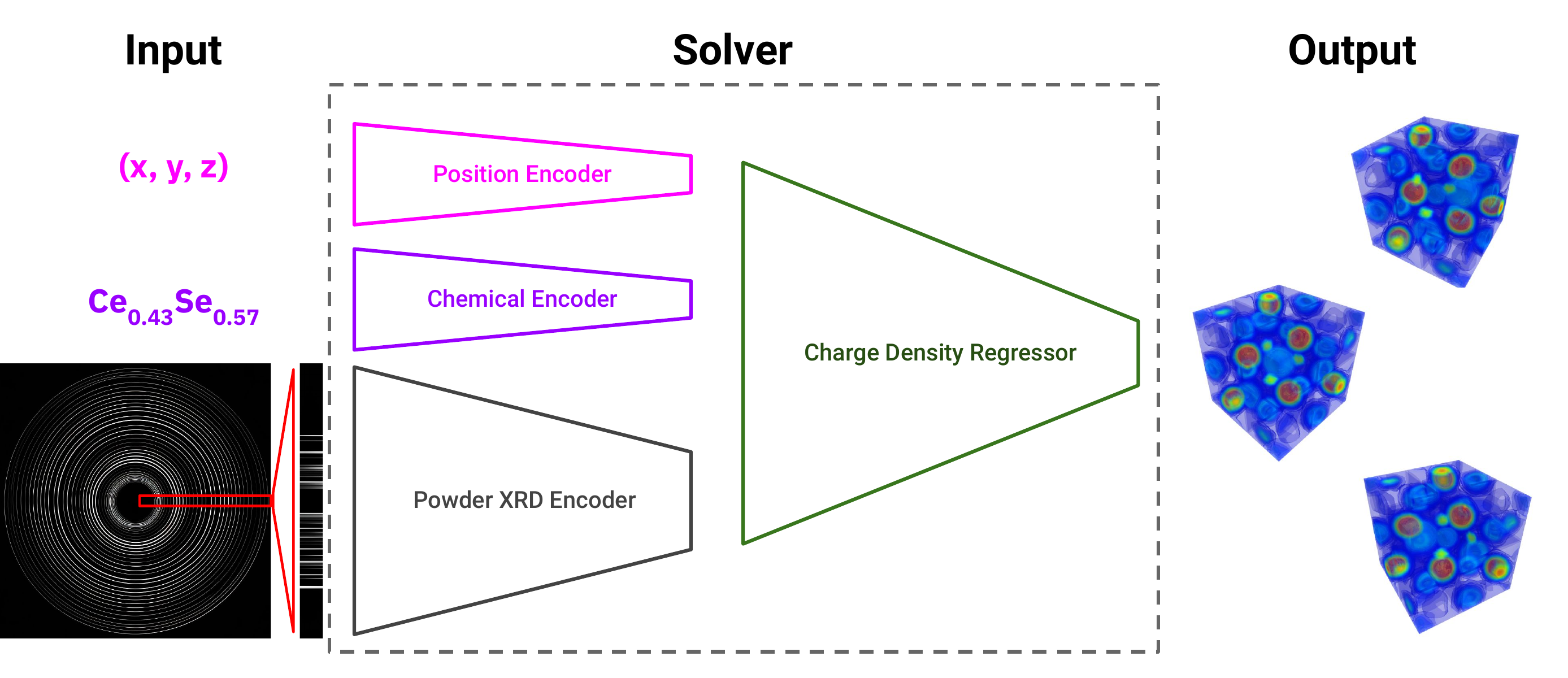}
    \caption{\textbf{\textit{CrystalNet} System Overview:} We use a multi-branch conditional coordinate-based representation to predict 3D transformed charge density maps, given powder x-ray diffraction patterns and chemical composition ratios.%\sjb{the model doesn't take a 2D detector image but a 1D pattern, right?  So it is a bit confusing to here show a 2D detector image.  Replace with an example of a 1D diffraction pattern?}
    }
    \label{fig:system_diagram}
\end{figure}

\section*{Results}

\subsection*{Generative Modeling}

\begin{figure}
    \centering
    \includegraphics[width=\textwidth]{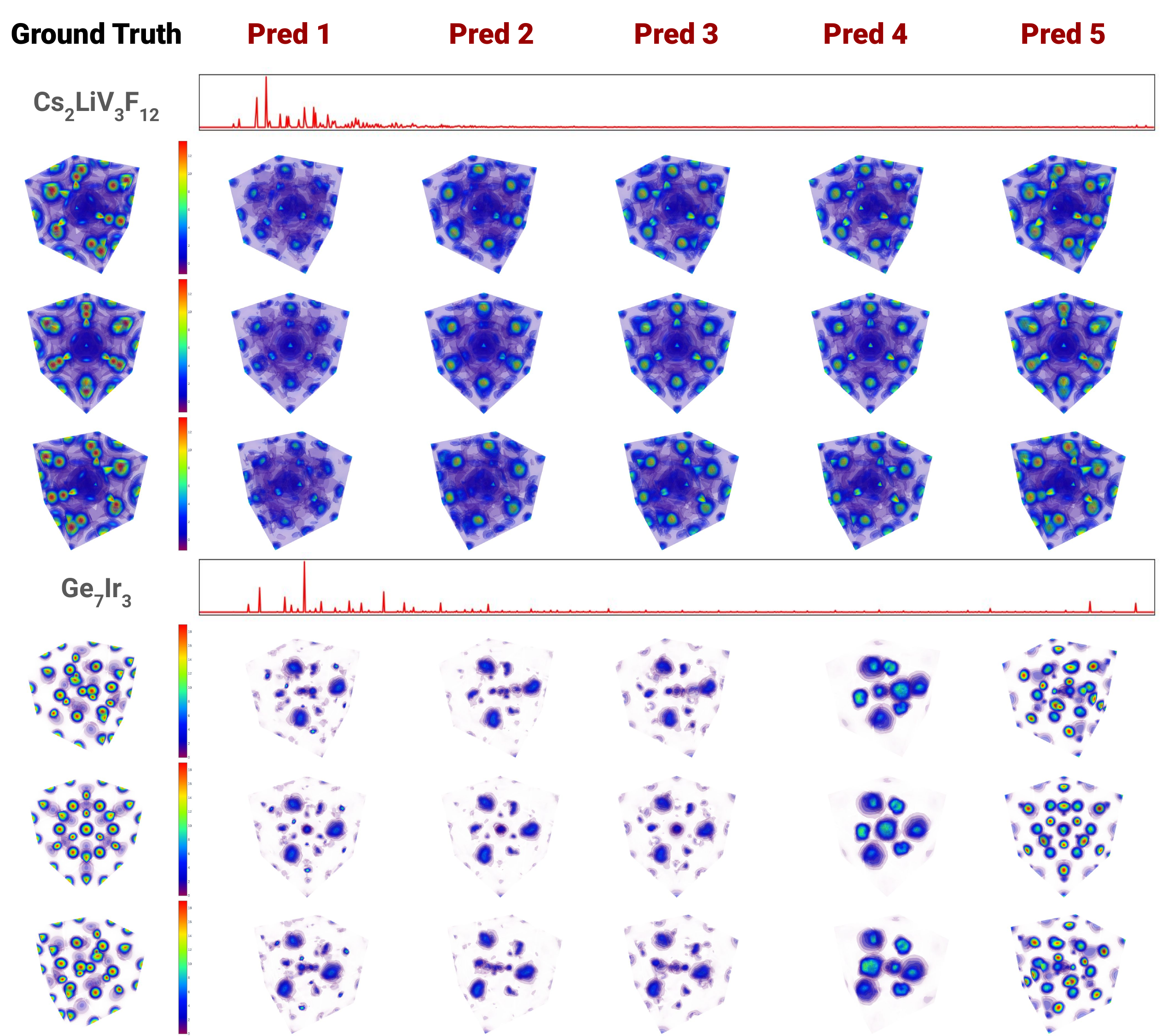}
    \caption{\textbf{Multi-View Variational Reconstructions:} We display CMED reconstructions of previously unseen crystals at multiple viewing angles. Given chemical composition and powder XRD as input, we can generate a distribution over latent codes, which we can then sample from to obtain multiple crystal reconstructions. This allows us to obtain better guesses, if the first prediction is sub-optimal. Even for failure cases like $Ge_{7}Ir_{3}$, we still see that sampling multiple times allows us to get a prediction that is closer to the ground truth. %\textcolor{red}{prediction with metrics on the side}
    }
    \label{fig:multi_angle_variational}
\end{figure}

Table \ref{tab:reconstruction_stats} shows generative modeling success metrics (SSIM, PSNR) on the cubic and trigonal crystal systems from powder XRD and chemical formula information. SSIM stands for structural similarity index, which measures the patchwise correspondence between two signals on a scale of $0$ (worst) to $1$ (best) \cite{ssim_wang2004image}. PSNR stands for peak signal-to-noise ratio, which measures the magnitude of the predicted charge density signal relative to the size of the errors in the prediction,
%\sjb{I am not sure I know what this means, so others in the crystallography community may not also.  Please briefly define/explain here} 
where higher values are better, and $\infty$ indicates perfect reconstruction \cite{psnr_hore2010image}. More details are available in Methods.

To demonstrate the functionality of our methodology, Figure \ref{fig:multi_angle_variational} shows sampled reconstructions of two testing crystals viewed from various angles, given only chemical composition and powder XRD as input. Inspired by variational approaches, we achieve multiple reconstructions by sampling from the conditional latent distribution \cite{kingma2013auto_vae}. We see that this stochasticity in output can be helpful if the initial guess is incorrect; in principle, we can resample to obtain a more reasonable prediction that matches the given XRD, as measured by the analytically solved forward process. 

\subsubsection*{Cubic System}

\begin{figure}
    \begin{subfigure}[b]{0.5\textwidth}
        \centering
        \includegraphics[width=\linewidth]{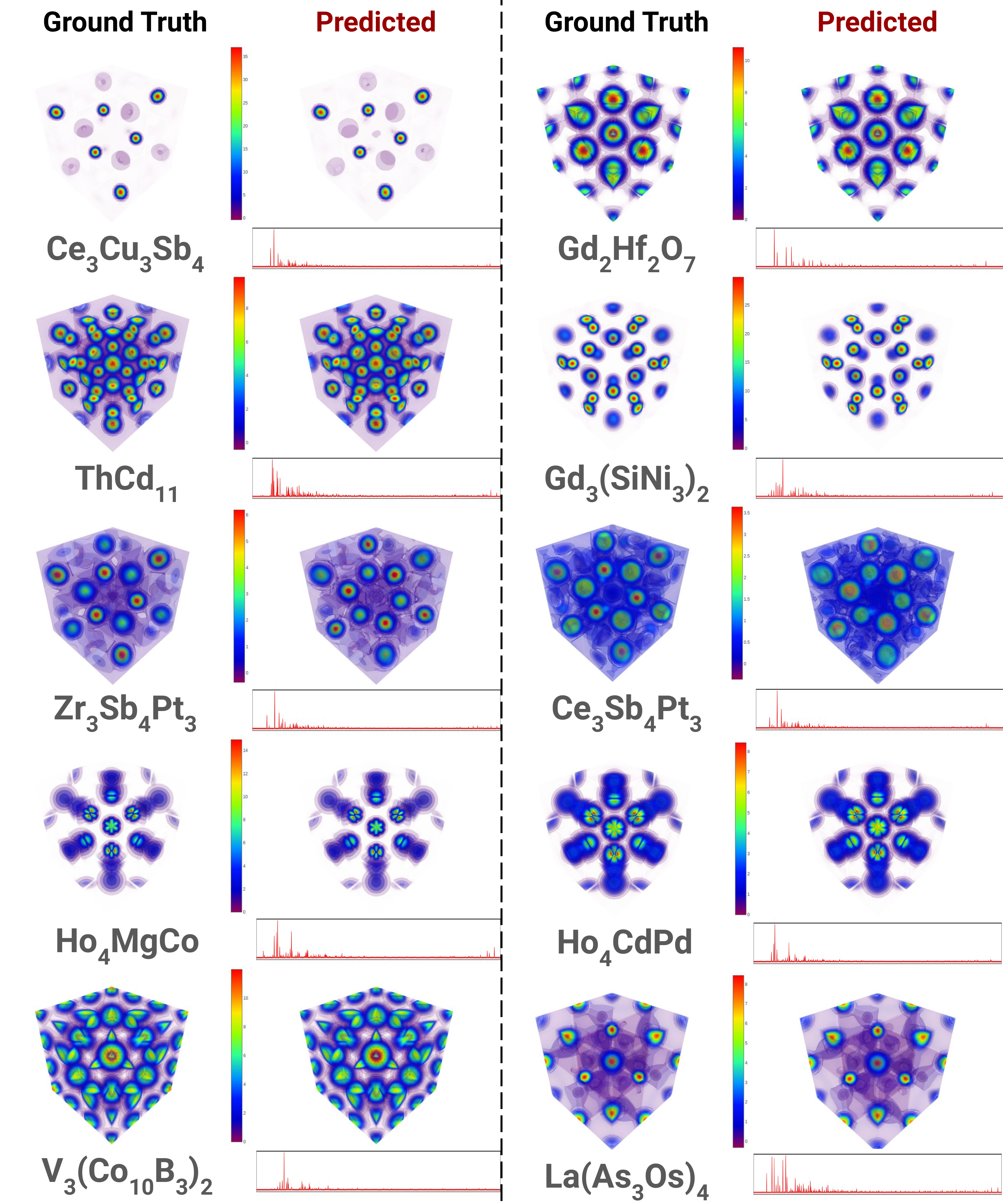}
        %\caption{\textbf{Cubic System Successes:} Alternating left-to-right: ground truth, corresponding \textit{CrystalNet} prediction.}
        \caption{}
        \label{fig:cubic_success}
    \end{subfigure}
    \begin{subfigure}[b]{0.5\textwidth}
        \centering
        \includegraphics[width=\linewidth]{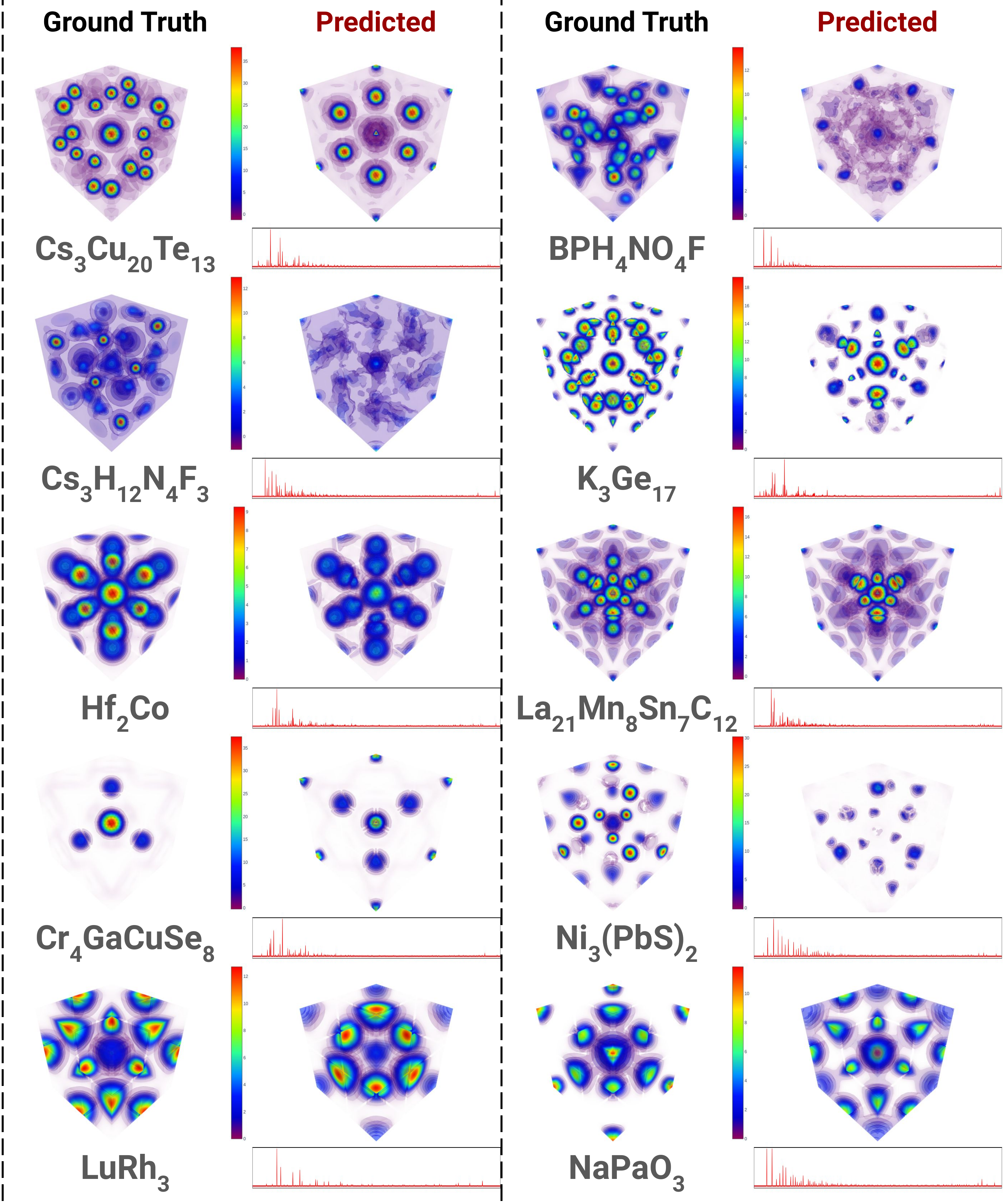}
        %\caption{\textbf{Cubic System Failures:} Alternating left-to-right: ground truth, corresponding \textit{CrystalNet} prediction.}
        \caption{}
        \label{fig:cubic_failure}
    \end{subfigure}
    \caption{\textbf{Cubic System Reconstructions.} Panel \ref{fig:cubic_success} shows the CMED plots for the success cases. Panel \ref{fig:cubic_failure} shows the failure cases. Ground truth and \textit{CrystalNet} prediction alternate left-to-right. Formulas are under ground truth images, and the corresponding powder XRD peak inputs are under predictions. Powder XRD peak inputs are visualized as relative intensity maps, with the diffraction angle (horizontal direction) increasing from $0^\circ \rightarrow 180^\circ$.  %\textcolor{red}{prediction with metrics on the side}
    }
    \label{fig:cubic_generation}
\end{figure}

See Figure \ref{fig:cubic_success} for success cases of cubic reconstruction, and Figure \ref{fig:cubic_failure} for failure cases of cubic reconstruction. 
Overall, reconstruction is very successful over a diverse range of crystal structures, judging from both visual and quantitative metrics. 
Even the cubic failure modes (Figure \ref{fig:cubic_failure}) still give good guesses for rough structural outlines, even if the details are slightly incorrect. 
We also observe, as expected, that crystals containing similar elements -- such as $Zr_{3}Sb_{4}Pt_{3}$ and $Ce_{3}Sb_{4}Pt_{3}$ -- have similar structures, albeit with different average charge densities. %\textcolor{red}{percentage of good predictions}

\subsubsection*{Trigonal System}

\begin{figure}
    \begin{subfigure}[b]{0.5\textwidth}
        \centering
        \includegraphics[width=\linewidth]{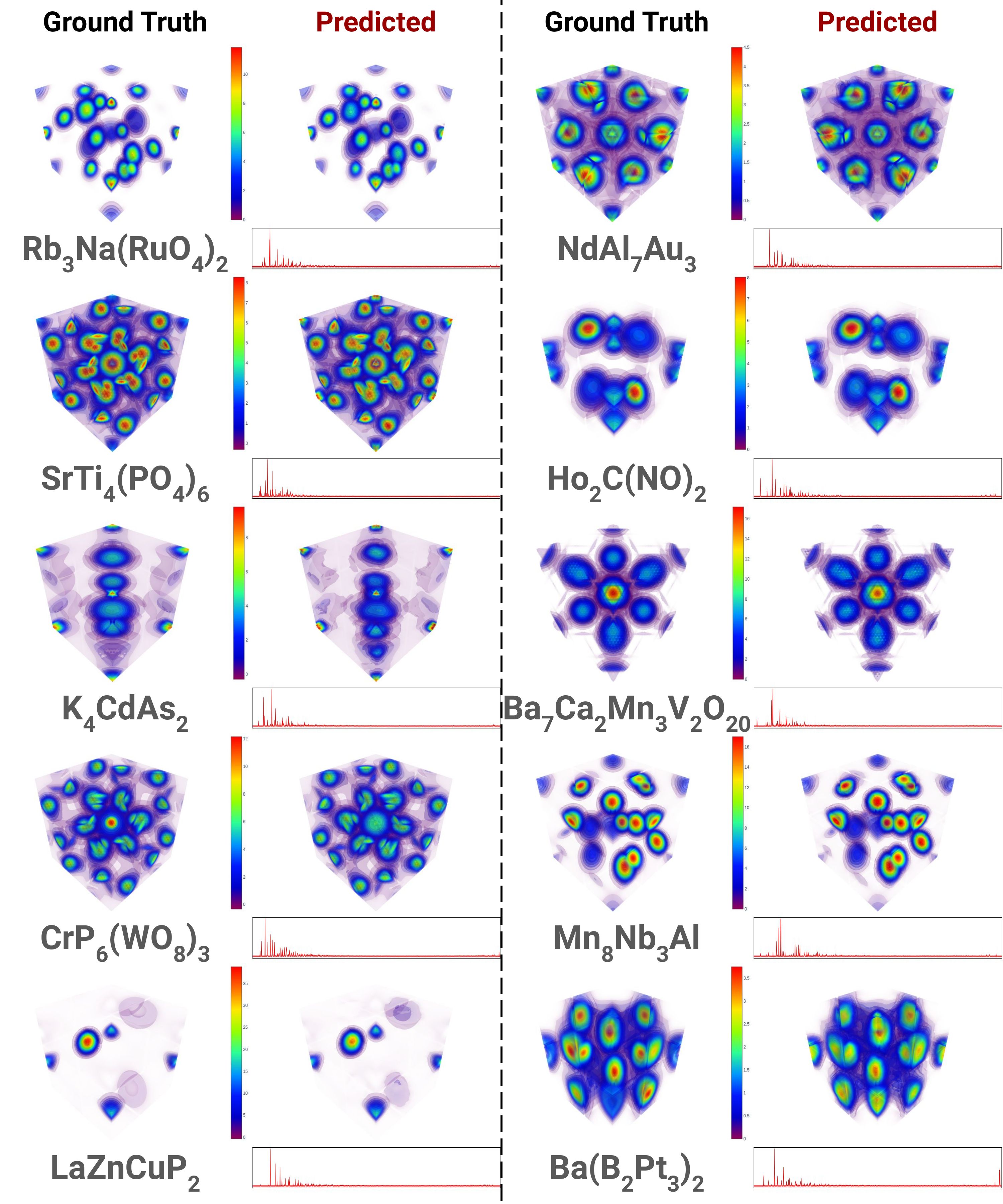}
        %\caption{\textbf{Trigonal System Successes:} Alternating left-to-right: ground truth, corresponding \textit{CrystalNet} prediction.}
        \caption{}
        \label{fig:trigonal_success}
    \end{subfigure}
    \begin{subfigure}[b]{0.5\textwidth}
        \centering
        \includegraphics[width=\linewidth]{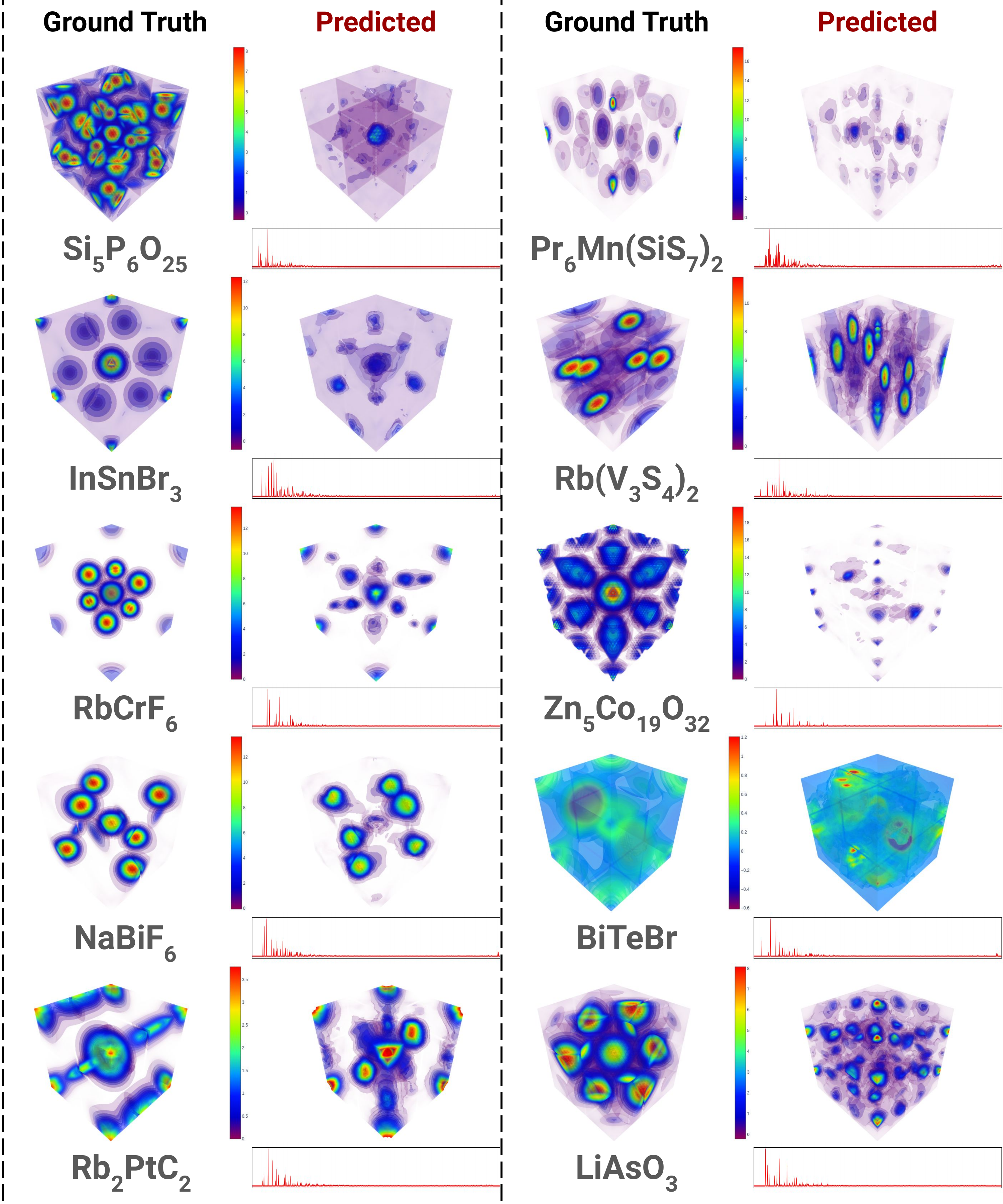}
        %\caption{\textbf{Trigonal System Failures:} Alternating left-to-right: ground truth, corresponding \textit{CrystalNet} prediction.}
        \caption{}
        \label{fig:trigonal_failure}
    \end{subfigure}
    \caption{\textbf{Trigonal System Reconstructions.} Panel \ref{fig:trigonal_success} shows the CMED plots for the success cases. Panel \ref{fig:trigonal_failure} shows the failure cases. Ground truth and \textit{CrystalNet} prediction alternate left-to-right. Formulas are under ground truth images, and the corresponding powder XRD peak inputs are under predictions. Powder XRD peak inputs are visualized as relative intensity maps, with the diffraction angle (horizontal direction) increasing from $0^\circ \rightarrow 180^\circ$. %\textcolor{red}{prediction with metrics on the side}
    }
    \label{fig:trigonal_generation}
\end{figure}

See Figure \ref{fig:trigonal_success} for success cases of trigonal reconstruction, and Figure \ref{fig:trigonal_failure} for failure cases of trigonal reconstruction. 
In the trigonal success cases (Figure \ref{fig:trigonal_success}), we see that the model is able to successfully solve crystal structures with considerably lower symmetry than the examples in the cubic system. 

We see that the failure cases (Figure \ref{fig:trigonal_failure}) are a bit more apparent for the trigonal system than for the cubic system. We believe this is due to crystals in the cubic system typically being fairly symmetric, and the consistent $90^\circ$ lattice angles (as opposed to the varying lattice angles in the trigonal system). Noticeably, the model seems to have difficulty predicting the high charge density regions. That being said, the failures still contain reasonable information about the structure, which can be used as a first step in an iterative structural refinement process. 

\begin{table}
    \centering
    \begin{tabular}{c | c c c}
    \textbf{System} & \textbf{\# Test} & \textbf{SSIM} $(\mu \pm \sigma)$ & \textbf{PSNR} $(\mu \pm \sigma)$ \\
    \hline
    Cubic & $500$ & $0.934 \pm 0.149$ & $43.0 \pm 12.7$ \\
    Trigonal & $237$ & $0.741 \pm 0.215$ & $27.8 \pm \text{\ \ } 8.1$ \\
    \end{tabular}
    \caption{\textbf{Reconstruction Performance} from combined powder XRD and chemical formula. SSIM ranges between $[0, 1]$, where higher is better. PSNR is unbounded, and higher values are better.} 
    \label{tab:reconstruction_stats}
\end{table}

\subsection*{Data Ablation}

We conduct ablation studies on the chemical formula information, since in reality, this data is known to varying degrees during the crystallographic process. We try three ablations: (1) Eliminate elemental ratio information, with a $1$ in the composition vector if the element is contained in the material, and $0$ otherwise; (2) Randomly drop one element from the ratio-free composition information, \textit{i.e.}, flip $1$ to $0$ for a singular randomly selected element (at least one element must be known, so we do not drop elements if the material contains a singular element); (3) No elemental information at all, leaving only XRD. In all these experiments, full XRD information was retained in all these ablation studies. See Table \ref{tab:ablation_stats} for the results of the ablation studies.

\begin{figure}
    \begin{subfigure}[b]{0.5\textwidth}
        \centering
        \includegraphics[width=\linewidth]{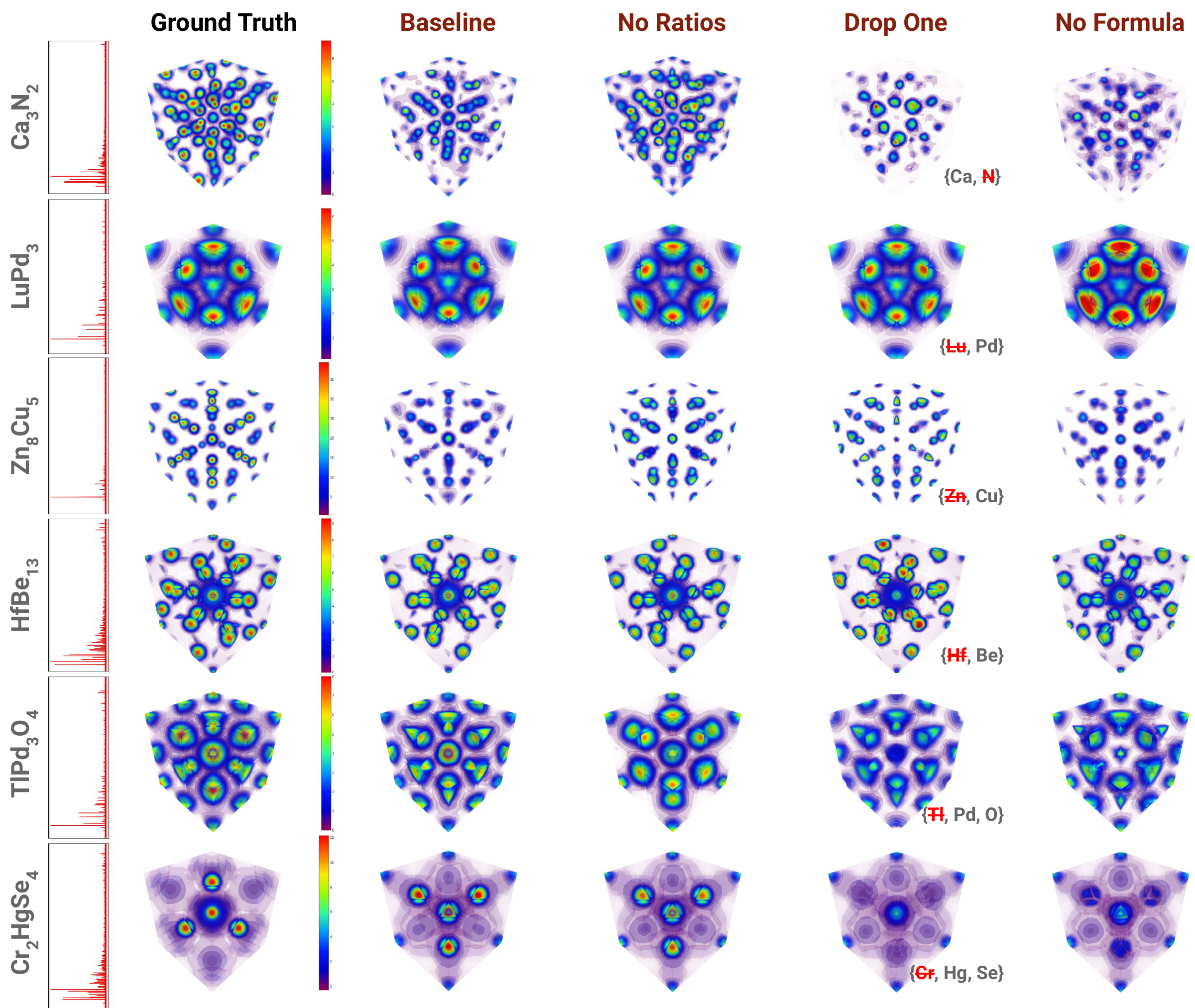}
        %\caption{\textbf{Cubic System Ablations:} Left-to-right: ground truth, full information (XRD + formula) prediction, excluding elemental ratios (XRD + elements contained) prediction, randomly dropped element (XRD + all but one elements contained) prediction, XRD-only prediction.}
        \caption{}
        \label{fig:cubic_ablation}        
    \end{subfigure}
    \begin{subfigure}[b]{0.5\textwidth}
        \centering
        \includegraphics[width=\linewidth]{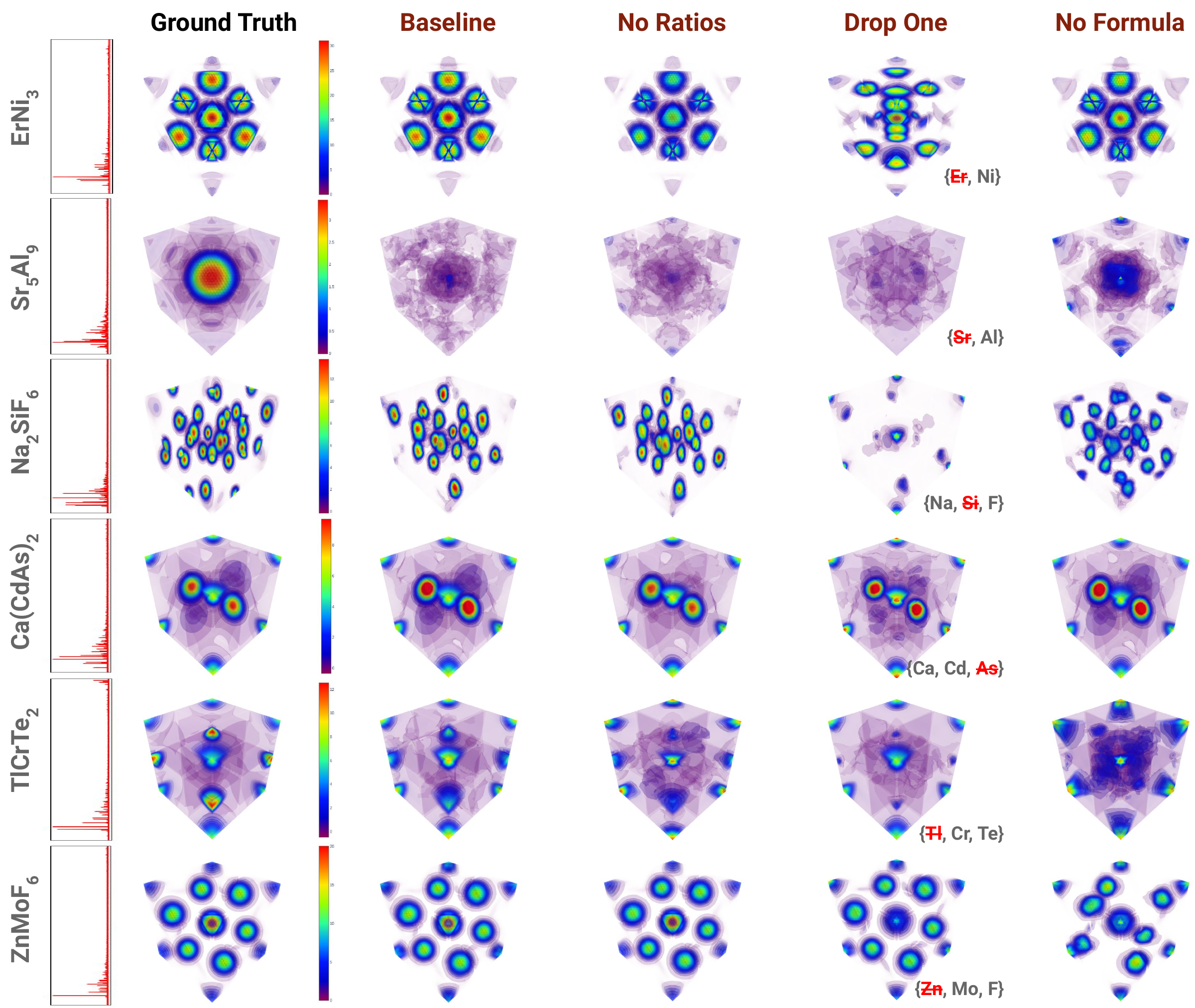}
        %\caption{\textbf{Trigonal System Ablations:} Left-to-right: ground truth, full information (XRD + formula) prediction, excluding elemental ratios (XRD + elements contained) prediction, randomly dropped element (XRD + all but one elements contained) prediction, XRD-only prediction.}
        \caption{}
        \label{fig:trigonal_ablation}
    \end{subfigure}
    \caption{\textbf{Ablation Studies:} Panel \ref{fig:cubic_ablation} shows cubic system ablations. Panel \ref{fig:trigonal_ablation} shows trigonal system ablations. Left-to-right in each panel: ground truth, full information (XRD + formula) prediction, excluding elemental ratios (XRD + elements contained) prediction, randomly dropped element (XRD + all but one element contained) prediction, XRD-only prediction. %\textcolor{red}{prediction with metrics on the side. We expect that the chemical formula only contributes to the absolute value of the prediction, while the structure information is encoded in XRD, as shown in a2/a4/a6. But AI learns some structure information (physics?) from chemical formulas like in a1/b1/b3/b6. Is this random? Maybe more tests on this later.}
    }
\end{figure}

\subsubsection*{Cubic System}

See Figure \ref{fig:cubic_ablation} for visualizations. As expected, as we ablate information about the chemical composition, the quantitative reconstruction performance, as measured by SSIM and PSNR, declines on the cubic system (Table \ref{tab:ablation_stats}). That being said, the visual and quantitative results indicate that even with heavy degradation in the elemental composition information inputted, we still achieve very reasonable reconstructions. This is likely because the cubic system has very regular structures with high symmetry priors, thus making it relatively easy to predict, even from just XRD data.

% \begin{figure}
%     \centering
%     \includegraphics[width=\linewidth]{results/ablation/Cubic Ablation.pdf}
%     \caption{\textbf{Cubic System Ablations:} Left-to-right: ground truth, full information (XRD + formula) prediction, excluding elemental ratios (XRD + elements contained) prediction, randomly dropped element (XRD + all but one elements contained) prediction, XRD-only prediction.}
%     \label{fig:cubic_ablation}
% \end{figure}

\subsubsection*{Trigonal System}

See Figure \ref{fig:trigonal_ablation} for visualizations. The trend of decreasing performance with decreasing degrees of chemical composition information still generally holds for the trigonal system (Table \ref{tab:ablation_stats}). However, different from the cubic system, removing the formula altogether from the trigonal reconstruction model's input does not have a significant performance difference (as measured by SSIM) from randomly dropping one element from the composition information. A reason for this anomaly may be that giving partial formula information may actually confuse the model more than giving no information at all, since there is considerably more variance in the trigonal as compared to the cubic system.

% \begin{figure}
%     \centering
%     \includegraphics[width=\linewidth]{results/ablation/Trigonal Ablation.pdf}
%     \caption{\textbf{Trigonal System Ablations:} Left-to-right: ground truth, full information (XRD + formula) prediction, excluding elemental ratios (XRD + elements contained) prediction, randomly dropped element (XRD + all but one elements contained) prediction, XRD-only prediction.}
%     \label{fig:trigonal_ablation}
% \end{figure}

\begin{table}
    \centering
    \begin{tabular}{l | c c}
    \textbf{System} & \textbf{SSIM} $(\mu \pm \sigma)$ & \textbf{PSNR} $(\mu \pm \sigma)$ \\
    \hline
    \rowcolor[gray]{0.95} \multicolumn{3}{l}{\textit{\textbf{Cubic}}} \\
    \hline
    \textit{Baseline} & $0.914 \pm 0.171$ & $40.9 \pm 12.8$\\
    \textit{No Ratios} & $0.916 \pm 0.161$ & $40.3 \pm 12.5$ \\
    \textit{Drop Element} & $0.887 \pm 0.170$ & $32.5 \pm \text{\ \ } 9.0$ \\
    \textit{No Formula} & $0.868 \pm 0.174$ & $30.0 \pm \text{\ \ } 8.6$\\
    \hline
    \rowcolor[gray]{0.95} \multicolumn{3}{l}{\textit{\textbf{Trigonal}}} \\
    \hline
    \textit{Baseline} & $0.732 \pm 0.207$ & $26.9 \pm \text{\ \ } 7.2$\\
    \textit{No Ratios} & $0.718 \pm 0.209$ & $26.8 \pm \text{\ \ } 7.2$\\
    \textit{Drop Element} & $0.695 \pm 0.203$ & $25.4 \pm \text{\ \ } 6.1$ \\
    \textit{No Formula} & $0.703 \pm 0.207$ & $25.5 \pm \text{\ \ } 6.1$\\
    \end{tabular}
    \caption{\textbf{Ablation Performance} at various levels of input information: (1) \textit{Baseline}: Powder XRD + full chemical composition information; (2) \textit{No Ratios}: Powder XRD + elements contained, without any information about their ratios; (3) \textit{Drop Element}: Powder XRD + elements contained, with one element randomly dropped; (4) \textit{No Formula}: Powder XRD information only.} 
    \label{tab:ablation_stats}
\end{table}

\section*{Discussion}

To the best of our knowledge, this is one of the world's first successful attempts at large-scale reconstruction of crystals in the cubic and trigonal systems. This is significant because it can pave the way for fully automated solutions to crystal structures from powder XRD data, potentially speeding up materials discovery and analysis by orders of magnitude. Furthermore, even if the structure initially predicted by our method is not correct, it can still be used as a first guess in the iterative refinement process, or we can even re-sample from the latent space to generate a new candidate (since we use a variational approach).

\subsection*{Limitations and Future Work}

All the experiments conducted were on simulated powder x-ray diffraction patterns. Furthermore, many of the materials in the Materials Project are theoretical materials that have never been synthesized \cite{materials_project_jain2013commentary}. This still provides us valid data pairs to train and evaluate our model, since generating XRD from crystal structure is an analytically solved problem \cite{crystallography_giacovazzo2002fundamentals}. However, this also means that much of the data is free from defects we would find in experimental data, \textit{e.g.}, peak broadening, missing peaks \cite{billinge2007problem, simon_powder_crystallography_dinnebier2008powder}. %Indeed, we modeled the powder XRD peaks as Dirac delta functions, rather than Gaussians \textcolor{red}{check}. 
Thus, while we have shown that deep learning methods, in principle, can work to solve the structure problem, there will still need to be future work to overcome this sim2real gap.

Furthermore, we solved only the two most symmetrical crystal systems, out of seven total \cite{crystallography_giacovazzo2002fundamentals}. Based on our preliminary explorations on the other five systems, we expect that this method, with appropriate tweaks, will be applicable to them. Yet, future exploration needs to be done to adapt our approach to these other systems. In particular, the decision to encode the input coordinates (see Methods) \cite{tancik2020fourier} with even and odd sinusoidal transformation functions may not translate so well to more asymmetrical molecules.

Additionally, solving crystal structures can be a one-to-many problem, in the cases of degraded XRD and/or chemical composition data. Although the variational approach allows us to have variation in the output via re-sampling from the latent space (see Methods), we seek more principled ways to model the uncertainty in our predictions.

Also, our representation of chemical composition information only tells the model which elements are contained, but it does not encode information about the chemical properties. In future works, we can perhaps incorporate some prior chemical knowledge, \textit{e.g.}, atomic mass, period, group.

%Finally, the output of the proposed system is an estimate of the charge distribution function in an arbitrary coordinate system (fractional coordinates along the $[0, 1]^3$ unit cell in the crystallographic basis), which is not a conventional solution format typically encountered in classic crystallography processes. We envision that additional machine learning processes can be trained to interpret the estimated charge distribution and express it in more conventional forms, such as a list of atoms and physical coordinates. These post-solution interpretive steps would involve the use of established segmentation and recognition algorithms that would identify and estimate the locations of discrete particles \cite{hoffmannDataDrivenApproachEncoding2019}. Similarly, additional processes would need to linearly transform the coordinate space from fractional coordinates in the crystallographic basis to a real-space Cartesian basis, involving the knowledge or prediction of lattice parameters \cite{lattice_prediction_from_xrd_chitturi2021automated}.

\section*{Methods}

\subsection*{Dataset}

\subsubsection*{Materials Project}
We get our data from the Materials Project \cite{materials_project_jain2013commentary}, which has publicly available standard data on over 150,000 inorganic compounds, largely for materials in the Inorganic Crystal Structure Database (ICSD) \cite{icsd_belsky2002new}. Some of the material properties are experimentally observed, while others are calculated with Density Functional Theory (DFT) \cite{vasp_hafner2008ab, materials_project_extra_info}.

\subsubsection*{Data Cleaning}

We ensure there is no train-test leakage in the dataset, as follows. Our criteria for whether two molecules are ``duplicates'' is that they have the same (1) chemical formula; \textit{and} (2) spacegroup. We go through our datasets and find all the molecules that have the same formula-spacegroup combination. Out of the molecules that share the same formula-spacegroup combination, we remove all but one of them from our dataset. 

\subsubsection*{Crystal Systems}

We use data from the cubic and trigonal crystal systems, which constitute two out of the seven total crystal systems. We only experiment on these two systems in this preliminary study because the intra-crystal axial lengths are equal (\textit{i.e.}, $a = b = c$), %are well-suited to our representation of the electron density map as a unit cell (\textit{i.e.}, $0 \leq x, y, z \leq 1$, where $x, y, z$ are on the same absolute scale). Since the intra-crystal axial lengths are the same, 
which eliminates the need to predict the axial lengths (whether implicitly as an intermediate calculation, or explicitly as the model's output), and allows us to focus on predicting charge densities. %In other crystal systems, the mismatch between axial lengths may mean that a change of $k\%$ along the $x$ axis is different than a change of $k\%$ along the $y$ axis. 
See Table \ref{tab:dataset_size} for the numbers of crystals used in our experiments.

\begin{table}[h]
    \centering
    \begin{tabular}{l|r r r}
        \textbf{Crystal System} & \textbf{Train} & \textbf{Val} & \textbf{Test} \\
        \hline
        \textbf{Trigonal} & $9177$ & $241$ & $237$ \\
        \textbf{Cubic} & $16378$ & $250$ & $500$ \\
    \end{tabular}
    \caption{\textbf{Number of crystals.} All samples are obtained from the Materials Project \cite{materials_project_jain2013commentary}. Training data consists of a mix of stable and unstable crystals, while validation and testing data is purely stable.}
    \label{tab:dataset_size}
\end{table}
% Trigonal: 
% - Training: 2094 stable, 8366 unstable
% - Validation: 245 stable
% - Testing: 241 stable

% Cubic:
% - Training: 5032 stable, 11527 unstable
% - Validation: 250 stable
% - Testing: 500 stable

\subsubsection*{X-Ray Diffraction Patterns}

We use the theoretically calculated powder x-ray diffraction patterns from the Materials Project API. The diffraction angle ranges used were between $0^\circ$ and $180^\circ$. More detail is available in the references \cite{diffraction_calculation_de2012structure, materials_project_diffraction_calculation}.

The simulated patterns are generated using the MoK$\alpha$ wavelength of $0.711$~\r{A}.
%\sjb{I think it was a mistake to use patterns in two-theta using Mo Kalpha.  In the next study we should switch everything to Q}
Depending on the atom types present in the compounds, the amplitude of the powder XRD patterns may vary drastically. This variation can be inherently problematic for most machine learning algorithms \cite{ioffe2015batchnorm}. To solve this issue, we normalize the peak intensities so that the highest peak intensity is set to $1$. While this normalization process does reduce some of the information related to specific atom species, it retains the relative differences between them. Consequently, when the chemical formula is provided, or even if only partial information about the atom species is available, we can still reconstruct the structure with the correct atom types.
 
% We use two wavelengths: MoK$\alpha$ ($0.711$ \r{A}) and CrK$\alpha$ ($2.291$ \r{A}). We have two, rather than one, for the following reasons: (1) we want to investigate whether more diffraction patterns enables better prediction; (2) we want to see the difference in predictive ability between longer (CrK$\alpha$) and shorter (MoK$\alpha$) wavelengths \footnote{We did not conduct our study on other wavelengths, because each XRD pattern must be simulated from scratch with the API, which is computationally prohibitive for the current stage of the study. We intend to investigate more wavelengths in future studies.}.

\subsubsection*{Chemical Composition}

We also incorporate the chemical composition, that is, the molar ratios of the elements contained in the material. We include this because chemical composition is often known, at least to some degree. We also test the robustness of the model by ablating this information to various degrees in our experiments.

\subsubsection*{Charge Density}

For the training, validation, and testing data, we use electron density maps from Materials Project DFT calculations \cite{materials_project_charge_density_calculation, vasp_chgcar, pyrho_shen2022representation}.  These are in a crystallographic basis, which depends discontinuously on the crystal system and details of the unit cell size and shape as we move from one material to another. 
We resample the electron densities within the unit cell onto a grid that has 50 voxels along each axis, with the locations of the voxels expressed in fractional coordinates.  We use PyRho (a library from the Materials Project) \cite{pyrho, pyrho_shen2022representation} to do this via Fourier interpolation.  The charge densities are further normalized to be expressed in $\frac{e^{-}}{\text{\r{A}}^{3}}$. This will give different spatial resolutions for different structures, but has the advantage that it gives a representation that is a uniformly shaped array for all materials. %\sjb{Gabe and Hod, please check this language is correct....do we use the charge density in the feature vector?}.

We call this quantity the Cartesian mapped electron density (CMED). 
The result of the normalization and resampling is a grid of $50\times 50\times 50$ voxels. For visualization we can project this onto a Cartesian coordinate system with orthonormal basis vectors.
The CMED is distorted from the real electron density by the procedure, but it allows us to visualize all structures, from all crystal systems and unit cells, on the same coordinate system.
However, more importantly, it allows in principle a single ML model to be trained on structures from all the different space groups and crystal systems.

To get from the CMED predicted by our model to an undistorted electron density the inverse mapping must be carried out.
If the unit cell of the unknown structure is indexed and the lattice parameters are known, this is straightforwardly done by plotting the voxels in the same order in the new basis.

In practice, we seek an end-to-end procedure that can discover the unit cell parameters as part of the automated process.
This has not been done in the current paper, but we believe it will be straightforward. 
Indeed, there is already evidence that such information can be obtained straightforwardly by ML \cite{lattice_prediction_from_xrd_chitturi2021automated,guccioneExtractionCrystalCell2023,garcia2019learning}.

\subsection*{Neural Network Design}

\begin{figure}
    \centering
    \includegraphics[width=\linewidth]{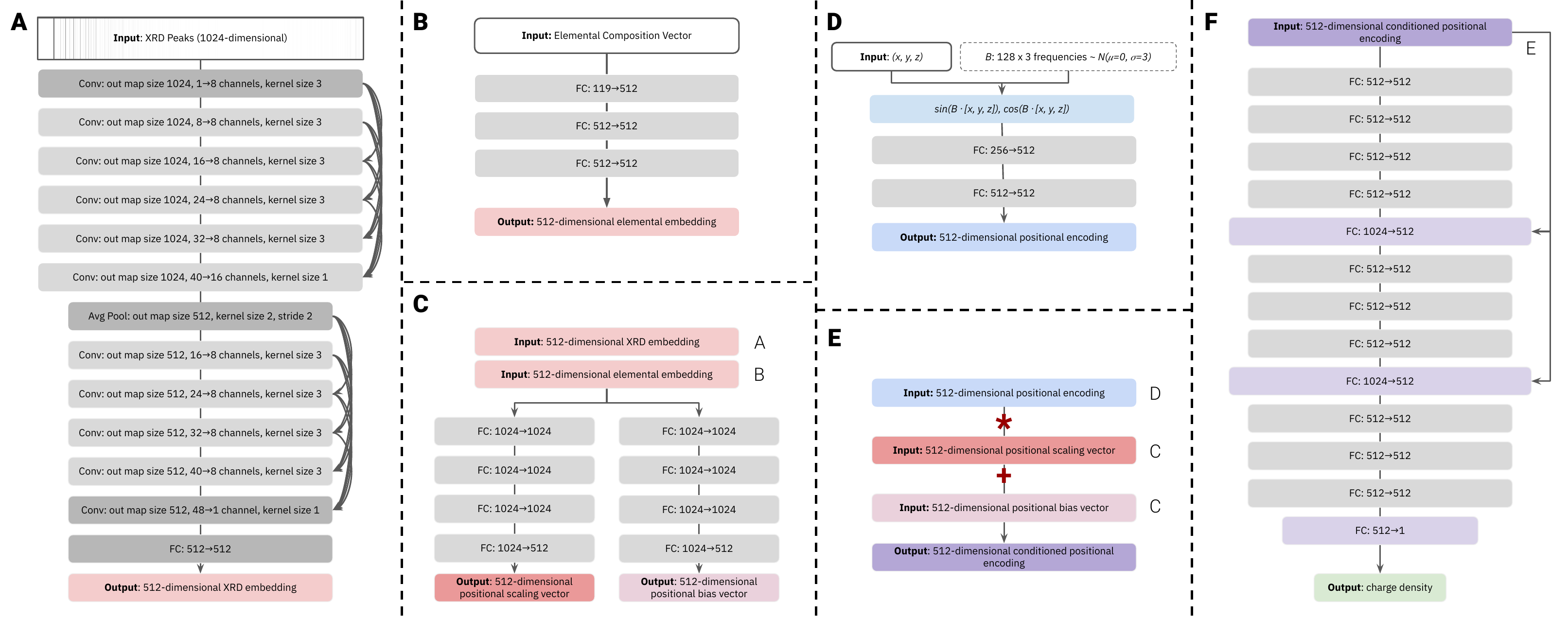}
    \caption{\textbf{\textit{CrystalNet} Architecture:} Panel A shows the architecture of the x-ray diffraction encoder, Panel B shows the architecture of the elemental composition encoder, Panel C shows the architecture of the feature fusion network, Panel D shows the architecture of the positional encoder, Panel E shows the architecture of the conditioning network, and Panel F shows the architecture of the final charge density regressor.}
    \label{fig:network_architecture}
\end{figure}

\begin{figure}
    \centering
    \includegraphics[width=\linewidth]{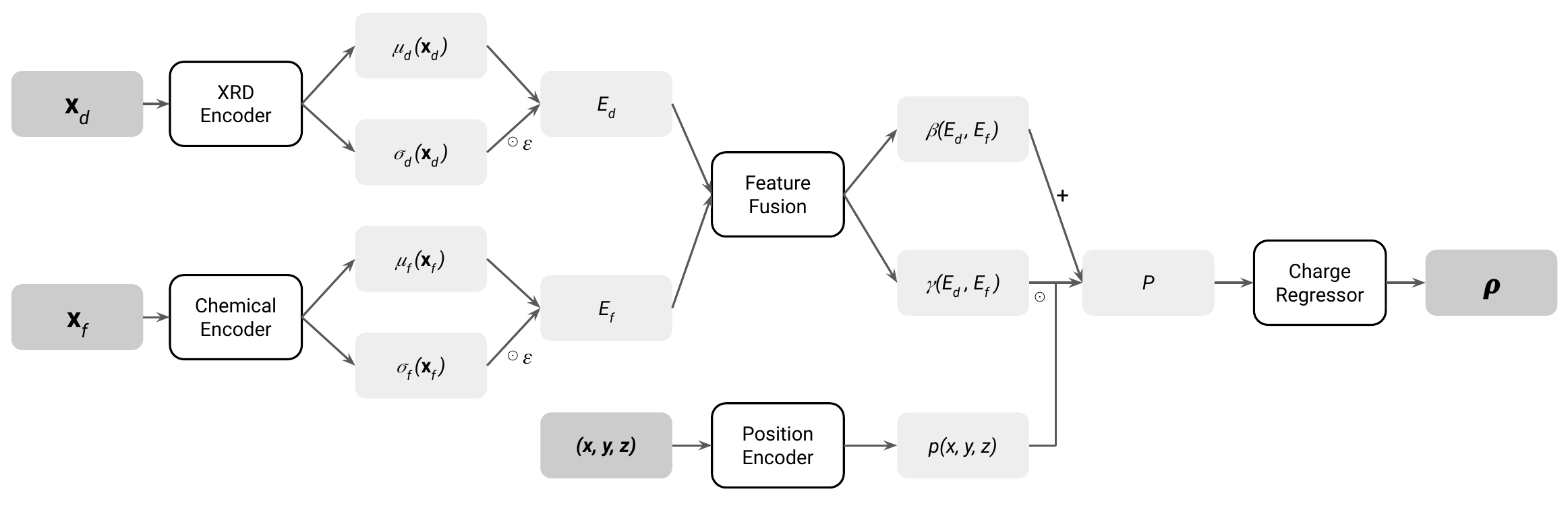}
    \caption{\textbf{\textit{CrystalNet} System Diagram:} XRD and chemical input are passed through encoders, which output latent feature distributions. Sampling from those latent distributions, we fuse the features with the positional $(x, y, z)$ query, and pass the fused information through the charge density regressor to get our final charge density prediction.}
    \label{fig:processing_steps}
\end{figure}

See Figure \ref{fig:network_architecture} for the layer-by-layer neural network architecture. See Figure \ref{fig:processing_steps} for a mid-level system diagram that shows how the components interact.


\subsubsection*{Variational Approach}

We adopt a variational approach \cite{kingma2013auto_vae, higgins2016beta_vae} for powder XRD and formula embedding prediction. Particularly, rather than deterministically predicting the embeddings, we predict the means and standard deviations of the embedding distributions, which are modeled as multivariate Gaussian distributions. Thus, we have
\begin{equation}\label{eqn:variational}
    E(\mathbf{x}) \sim \mathcal{N}(\mu(\mathbf{x}), \sigma^2(\mathbf{x}))
\end{equation}
where $E$ is a sample from the distribution of formula- or XRD-conditioned embeddings, $\mathbf{x}$ is the corresponding formula or XRD input, $\mu$ is the neural network function that regresses the mean, and $\sigma$ is the neural network function that regresses the standard deviation. We use the reparameterization
\begin{equation}\label{eqn:reparam}
    E(\mathbf{x}) = \mu(\mathbf{x}) + \epsilon * \sigma(\mathbf{x})
\end{equation}
where $\epsilon$ is unit Gaussian noise, to make the process differentiable \cite{kingma2013auto_vae}.

The justification for doing this is as follows: (1) Crystallographic inference, \textit{i.e.}, predicting molecular structure given XRD and formula, can be a one-to-many problem, so a non-deterministic approach is appropriate for modeling these multiple outputs. (2) Crystallography is an iterative design process. The variational approach allows us to resample candidate structures, if the first prediction is not appropriate. (3) Variational approaches allow the model to learn a smoother latent space, which may generalize better to out-of-training-distribution inputs \cite{kingma2013auto_vae}.

We note that although Figure \ref{fig:network_architecture} depicts the powder XRD (Panel A) and formula (Panel B) encoders as deterministic networks, this is only for the sake of simplicity in the illustration. In reality, we have two versions of each network, one for regressing $\mu$, and the other for regressing $\sigma$, which are then combined to produce the actual embedding, according to Equation \ref{eqn:reparam}.

\subsubsection*{XRD Encoder}

The powder XRD encoder is shown in Figure \ref{fig:network_architecture}, Panel A. The inputs are the extracted peaks $\mathbf{x_{d}}$ from the x-ray diffraction patterns, which are normalized such that the highest peak is at intensity $1$. They are represented as vectors with with 1024 pixels of resolution, where the value at each pixel represents the intensity of the diffraction pattern at that location. The outputs are 512-dimensional embeddings $E_{d}(\mathbf{x_{d}})$. 

The architecture is an adaptation of the DenseNet architecture for vector (rather than image matrix) inputs, with the most important design characteristic being the densely connected concatenations between convolutional feature maps \cite{densenet_huang2017densely}. Every convolutional layer (except the last one) is followed by LayerNorm \cite{ba2016layernorm} and ReLU; the final linear layer is followed by BatchNorm \cite{ioffe2015batchnorm}. 
We re-emphasize that technically, we have two versions of the XRD encoder under our variational framework: one for regressing $\mu_{d}(\mathbf{x_{d}})$, and one for regressing $\sigma_{d}(\mathbf{x_{d}})$, to construct $E_{d}(\mathbf{x_{d}})$ as defined in Equation \ref{eqn:reparam}.

\subsubsection*{Formula Encoder}

The formula encoder is shown in Figure \ref{fig:network_architecture}, Panel B. The input is the empirical formula, represented as a 118-dimensional vector $\mathbf{x_{f}}$, where each index of the vector refers to the normalized amount (as defined by number of atoms) of the element with that atomic number that is contained in the formula. (For instance, if the formula was $H_{2}O$, we would first normalize that to $H_{0.66}O_{0.33}$. The resultant vector would contain $0.66$ at index $1$, the atomic number of hydrogen; $0.33$ at index $8$, the atomic number of oxygen; and $0$ everywhere else.) The output is a 512-dimensional embedding $E_{f}(\mathbf{x_{f}})$.

The architecture is a simple MLP, in which every linear layer is followed by BatchNorm \cite{ioffe2015batchnorm} and ReLU. The only exception is that the last layer does not use ReLU.
We reiterate that we use a variational framework for regressing $E_{f}(\mathbf{x_{f}})$, which technically necessitates two versions of the encoder, one for $\mu_{f}(\mathbf{x_{f}})$, and one for $\sigma_{f}(\mathbf{x_{f}})$.

\subsubsection*{Feature Fusion Network}

The feature fusion network is shown in Figure \ref{fig:network_architecture}, Panel C. The inputs are the concatenated embeddings from the XRD encoders and formula encoders, such that we have a 1024-dimensional combined embedding. This combined embedding then gets passed through two MLPs with four linear layers each, and BatchNorm \cite{ioffe2015batchnorm} and ReLU following every linear layer. The outputs are two 512-dimensional embeddings, one for multiplicative interactions (labeled $\gamma(E_{d}, E_{f})$), the other for additive interactions (labeled $\beta(E_{d}, E_{f})$) with the positional encoding (described in next section).

\subsubsection*{Positional Encoder}

The positional encoder is shown in Figure \ref{fig:network_architecture}, Panel D. It takes in the $(x, y, z)$ coordinates as input. The inputted coordinates are normalized and centered, such that $-0.5 \leq x, y, z \leq +0.5$. The output is a 512-dimensional positional embedding. 

To process the input, we use modified random Fourier features \cite{tancik2020fourier}, according to the formula:
\begin{equation}\label{eqn:positional} 
    p([x, y, z]) = \text{MLP}([sin(B \cdot [x, y, z]^T), cos(B \cdot [x, y, z]^T)])
\end{equation}
We generate the frequency matrix $B \in \mathbb{R}^{m \times 3}$, where each $B_{ij} \sim \mathcal{N}(0, \sigma^2)$. (We set $m = 128, \sigma = 3$.) Then, we calculate $x' = B \cdot [x, y, z]^\text{T}$, which represents linear combinations of each of the coordinates. Then, we calculate $\text{sin}(x')$ and $\text{cos}(x')$ and concatenate them to get a $2m$-dimensional psuedo-Fourier series representation. We use this coordinate transformation for two reasons: (1) it approximates a high-dimensional Fourier series of the charge density map, which allows the model to capture high-frequency features (shown via Neural Tangent Kernel theory \cite{jacot2018ntk, tancik2020fourier}); (2) the cosine (periodic even function) and sin (periodic odd function) parameterizations allow us to encode the many inherent symmetries \cite{crystallography_giacovazzo2002fundamentals} in crystals. Finally, we pass the $2m$-dimensional pseudo-Fourier series representation through two linear layers, with a BatchNorm \cite{ioffe2015batchnorm} and ReLU in between; to get our positional encoding $p([x, y, z])$. 

\subsubsection*{Feature Conditioning}

The feature conditioner is shown in Figure \ref{fig:network_architecture}, Panel E. It takes as input the multiplicative embeddings $\gamma(E_d, E_f)$, additive embeddings $\beta(E_d, E_f)$, and positional encoding $p([x, y, z])$. It outputs $P$, the 512-dimensional feature-conditioned positional encoding.

The feature-conditioned positional encoding $P$ is calculated as:
\begin{equation} \label{eqn:film_conditioned}
    P = \gamma(E_d, E_f) * p([x, y, z]) + \beta(E_d, E_f)
\end{equation}
This is known as the feature-wise linear modulation (FiLM) \cite{perez2018film}, and is effective because it allows us to have both multiplicative and additive interactions during feature conditioning.

\subsubsection*{Charge Density Regressor}

The charge density regressor is shown in Figure \ref{fig:network_architecture}, Panel F. The input is $P$, the feature-conditioned positional encoding. The output is the charge density at the corresponding $(x, y, z)$ coordinates that $P$ was generated from.

The architecture is a MLP with BatchNorm \cite{ioffe2015batchnorm} and ReLU after every layer, except for the final layer. It also uses skip connections to encourage feature reuse, inspired by DeepSDF \cite{park2019deepsdf} and NeRF \cite{mildenhall2021nerf}.

\subsection*{Training Process}

\subsubsection*{Loss Function}

We minimize L1 Loss on the predicted charge densities, averaged over the entire batch:
\begin{equation}\label{eqn:l1_loss}
    L_1(\rho_\text{pred}, \rho_\text{gt}) = |\rho_\text{pred} - \rho_\text{gt}|
\end{equation}
Minimizing this loss encourages the predicted output to match the ground truth output. We call this the reconstruction loss. 

We also simultaneously minimize a KL-Divergence Loss on the predicted mean $\mu_{d}(\mathbf{x}_{d}), \mu_{f}(\mathbf{x}_{f})$ and standard deviation $\sigma_{d}(\mathbf{x}_{d}), \sigma_{f}(\mathbf{x}_{f})$ of the distribution of embedding vectors $E_{d}(\mathbf{x}_{d}), E_{f}(\mathbf{x}_{f})$ \cite{kl_div_kullback1951information, kingma2013auto_vae, higgins2016beta_vae}, similar to that in $\beta$-VAE \cite{higgins2016beta_vae}:
\begin{equation}\label{eqn:kl_loss}
    \beta L_{KL}(\mu(\mathbf{x}), \sigma(\mathbf{x})) = \frac{\beta}{N} \sum_{i=1}^{N} (1 + \text{log}(\sigma_{i}(\mathbf{x})^2) - \mu_{i}(\mathbf{x})^2 - \sigma_{i}(\mathbf{x})^2) \propto D_{KL}(q_\phi(E(\mathbf{x}) | \mathbf{x}) || p(E(\mathbf{x}))) 
\end{equation}
where $E(\mathbf{x}) = \mu(\mathbf{x}) + \epsilon * \sigma(\mathbf{x})$, $N = |E(\mathbf{x})| = |\mu(\mathbf{x})| = |\sigma(\mathbf{x})| = 512$ is the dimensionality of the embedding vector, $q_\phi(E(\mathbf{x}) | \mathbf{x})$ is the conditional distribution of the embedding vectors given the inputted XRD pattern or chemical formula, and $\beta$ is a weighting parameter given to the loss. This closed form is possible because we parameterize $p(E(\mathbf{x}))$ as $\mathcal{N}(0, \mathbf{I})$, following Kingma and Welling, who include a derivation in their paper \cite{kingma2013auto_vae}.
Intuitively, minimizing this loss encourages the XRD and formula embedding vectors to match multivariate Gaussian distributions, which not only smoothens the latent space, but encourages variation in the outputs, such that we can conduct an iterative refinement process in this one-to-many problem.

Thus, the total loss to be minimized is the sum of Equations \ref{eqn:l1_loss} and \ref{eqn:kl_loss}:
\begin{equation}\label{eqn:total_loss}
    L = L_1(\rho_\text{pred}, \rho_\text{gt}) + \beta L_{KL}(\mu_{d}(\mathbf{x}_{d}), \sigma_{d}(\mathbf{x}_{d})) + \beta L_{KL}(\mu_{f}(\mathbf{x}_{f}), \sigma_{f}(\mathbf{x}_{f}))
\end{equation}
The $\beta$ term tweaks the a balance between the reconstruction and the KL terms. Empirically, we set $\beta = 0.05$. 

\subsubsection*{Optimization Loop}

We train our model to minimize the total loss from Equation \ref{eqn:total_loss} for $1500$ epochs, with $128$ crystals per batch at a resolution of $10^3$ sampled charge densities per crystal. The charge densities are sampled via stratified bin sampling, where $x, y, z \sim \mathcal{U} [\frac{i}{S}, \frac{i + 1}{S}]$ (we set $S = 10$) -- this probabilistically allows us, over the course of the optimization procedure, to capture fine-resolution details of the electron density field, despite processor memory limits for individual batches \cite{mildenhall2021nerf}. 

We use the Adam \cite{kingma2014adam} optimizer. We follow a cosine annealing schedule with warm restarts \cite{loshchilov2016sgdr_cosine_warm_restart}, in which the learning rate decays from $10^{-3}$ to $10^{-6}$, then increases back to $10^{-3}$ and decays again to $10^{-6}$ over another cycle that has double the number of epochs: this helps the optimization procedure break out of local minima. The initial cycle length is $100$ epochs, and increases to $200, 400, \text{and } 800$ on the subsequent cycles, to constitute the $1500$ total epochs.

As data augmentation, we randomly add small Gaussian perturbations from $\mathcal{N}(0, 0.001^{2}\mathbf{I})$ to the inputted XRDs and chemical formula ratios (the perturbed input undergoes a ReLU, since we cannot have negative peaks or ratios). We also randomly shift the XRD patterns by less than $0.6^\circ$.

We save the version of the model that has the highest SSIM \cite{ssim_wang2004image} score on the validation set at the end of each epoch, where the model is given two guesses for each structure, and the rotation ($24$ ways) of the predicted structure that gives the highest SSIM score with the ground truth is used.

\subsection*{Evaluation Setup}

\subsubsection*{Procedure}

We run through the testing dataset, and give the model $5$ tries (via sampling from the latent space in the variational framework) to predict each crystal structure. (We give the model multiple tries because crystallography is typically an iterative refinement process, so we consider our model successful if it can give \textit{a} good guess.) For each guess, we rotate the predicted crystal $24$ ways (in multiples of $90^\circ$ about the $x, y, z$ unit cell axes) and take the best SSIM \cite{ssim_wang2004image} and PSNR \cite{psnr_hore2010image} over all these rotations, as compared to the ground truth crystal. Finally, we report the best results over all rotations of all guesses of each crystal structure. 

\subsubsection*{Metrics}

At evaluation time, we sample evenly to get a $50 \times 50 \times 50$ charge density map (\textit{i.e.}, 3D grid) for each crystal. We then use 3D SSIM \cite{ssim_wang2004image} and PSNR \cite{psnr_hore2010image} as our evaluation metrics on the resultant 3D grid. 

SSIM ranges from $0$ to $1$, where higher is better. It compares the structural similarity of the ground truth charge density map with the predicted charge density map. SSIM is calculated over patches of the 3D structure with a sliding cubic window of side length $7$, and then averaged over all such patches. The patch-wise formula is:
\begin{equation}\label{eqn:ssim}
    SSIM(\mathbf{x}, \mathbf{y}) = \frac{(2\mu_{x}\mu_{y} + C_1)(2\sigma_{xy} + C_2)}{(\mu_{x}^{2} + \mu_{y}^{2} + C_1)(\sigma_{x}^{2} + \sigma_{y}^{2} + C_{2})}
\end{equation}
where $\mathbf{x}, \mathbf{y}$ are the spatially corresponding patches of the ground truth and predicted electron density maps, $\mu_{x}, \mu_{y}$ are the mean charge densities (\textit{i.e.}, intensity) in those patches, $\sigma_{x}, \sigma_{y}$ are the standard deviations of the charge densities (\textit{i.e.}, contrast) in those patches, $\sigma_{xy}$ is the covariance between the position-wise charge densities in those patches, and $C_1, C_2$ are small constants for numerical stability. 

PSNR stands for peak signal-to-noise ratio, where higher is better. Its value is theoretically infinite. PSNR is calculated as follows \cite{van2014scikit_image}:
\begin{equation}\label{eqn:psnr}
    PSNR(\mathbf{X}_{gt}, \mathbf{X}_{pred}) = 10 * \text{log}_{10}\left(\frac{(\text{max}(\mathbf{X}_{gt}) - \text{min}(\mathbf{X})_{gt})^2}{MSE(\mathbf{X}_{gt}, \mathbf{X}_{pred})}\right)
\end{equation}
where $\mathbf{X}_{gt}, \mathbf{X}_{pred}$ are the ground truth and predicted charge density maps, respectively.

\subsection*{Formula Ablation Experiment}

To reduce the computational burden of these formula ablation studies, we make a few modifications. In the optimization loop, we only train for $700$ total epochs, use $8^3$ samples per crystal, and have only $1$ sample from the latent space per validation crystal.
Additionally, in testing, we give the model $3$ tries (instead of $5$) to predict each crystal via variational sampling. 
We can make these modifications because the purpose of these ablation experiments is to compare the predictive ability of the model at varying degrees of chemical composition information, rather than to optimize the model to perfect predictive ability. 
For fair comparison, we also recalculate the baseline model performance according to these pared-down protocols.

\subsection*{Data Availability}

Please visit the Materials Project website \cite{materials_project_jain2013commentary} to obtain the data: \url{https://next-gen.materialsproject.org/materials}.

\subsection*{Code Availability}

Code is available at \url{https://github.com/gabeguo/deep-crystallography-public}. 

\bibliography{sample,bg-pdf-standards,billinge-group-bib}

\begin{thebibliography}{10}
\urlstyle{rm}
\expandafter\ifx\csname url\endcsname\relax
  \def\url#1{\texttt{#1}}\fi
\expandafter\ifx\csname urlprefix\endcsname\relax\def\urlprefix{URL }\fi
\expandafter\ifx\csname doiprefix\endcsname\relax\def\doiprefix{DOI: }\fi
\providecommand{\bibinfo}[2]{#2}
\providecommand{\eprint}[2][]{\url{#2}}

\bibitem{crystallography_giacovazzo2002fundamentals}
\bibinfo{author}{Giacovazzo, C.}
\newblock \emph{\bibinfo{title}{Fundamentals of crystallography}}, vol.~\bibinfo{volume}{7} (\bibinfo{publisher}{Oxford university press, USA}, \bibinfo{year}{2002}).

\bibitem{crystallography_hammond2015basics}
\bibinfo{author}{Hammond, C.}
\newblock \emph{\bibinfo{title}{The basics of crystallography and diffraction}}, vol.~\bibinfo{volume}{21} (\bibinfo{publisher}{International Union of Crystallography texts on crystallography}, \bibinfo{year}{2015}).

\bibitem{lipson1935crystal}
\bibinfo{author}{Lipson, H.} \& \bibinfo{author}{Beevers, C.}
\newblock \bibinfo{journal}{\bibinfo{title}{The crystal structure of the alums}}.
\newblock {\emph{\JournalTitle{Proceedings of the Royal Society of London. Series A-Mathematical and Physical Sciences}}} \textbf{\bibinfo{volume}{148}}, \bibinfo{pages}{664--680} (\bibinfo{year}{1935}).

\bibitem{simon_powder_crystallography_dinnebier2008powder}
\bibinfo{author}{Dinnebier, R.~E.} \& \bibinfo{author}{Billinge, S.~J.}
\newblock \emph{\bibinfo{title}{Powder diffraction: theory and practice}} (\bibinfo{publisher}{Royal society of chemistry}, \bibinfo{year}{2008}).

\bibitem{henry_lipson_daniel1943x}
\bibinfo{author}{Daniel, V.} \& \bibinfo{author}{Lipson, H.~S.}
\newblock \bibinfo{journal}{\bibinfo{title}{An x-ray study of the dissociation of an alloy of copper, iron and nickel}}.
\newblock {\emph{\JournalTitle{Proceedings of the Royal Society of London. Series A. Mathematical and Physical Sciences}}} \textbf{\bibinfo{volume}{181}}, \bibinfo{pages}{368--378} (\bibinfo{year}{1943}).

\bibitem{henry_lipson1942structure}
\bibinfo{author}{Lipson, H.~S.} \& \bibinfo{author}{Stokes, A.}
\newblock \bibinfo{journal}{\bibinfo{title}{The structure of graphite}}.
\newblock {\emph{\JournalTitle{Proceedings of the Royal Society of London. Series A. Mathematical and Physical Sciences}}} \textbf{\bibinfo{volume}{181}}, \bibinfo{pages}{101--105} (\bibinfo{year}{1942}).

\bibitem{lipson1984study}
\bibinfo{author}{Lipson, H.}
\newblock \emph{\bibinfo{title}{The study of metals and alloys by X-ray powder diffraction methods}} (\bibinfo{publisher}{University College Cardiff Press Cardiff}, \bibinfo{year}{1984}).

\bibitem{billinge2007problem}
\bibinfo{author}{Billinge, S.~J.} \& \bibinfo{author}{Levin, I.}
\newblock \bibinfo{journal}{\bibinfo{title}{The problem with determining atomic structure at the nanoscale}}.
\newblock {\emph{\JournalTitle{science}}} \textbf{\bibinfo{volume}{316}}, \bibinfo{pages}{561--565} (\bibinfo{year}{2007}).

\bibitem{davidStructureDeterminationPowder2008a}
\bibinfo{author}{David, W. I.~F.} \& \bibinfo{author}{Shankland, K.}
\newblock \bibinfo{journal}{\bibinfo{title}{Structure determination from powder diffraction data}}.
\newblock {\emph{\JournalTitle{Acta Crystallogr A Found Crystallogr}}} \textbf{\bibinfo{volume}{64}}, \bibinfo{pages}{52--64}, \doiprefix\url{10.1107/S0108767307064252} (\bibinfo{year}{2008}).

\bibitem{alphafold_jumper2021highly}
\bibinfo{author}{Jumper, J.} \emph{et~al.}
\newblock \bibinfo{journal}{\bibinfo{title}{Highly accurate protein structure prediction with alphafold}}.
\newblock {\emph{\JournalTitle{Nature}}} \textbf{\bibinfo{volume}{596}}, \bibinfo{pages}{583--589} (\bibinfo{year}{2021}).

\bibitem{alphafold2_bryant2022improved}
\bibinfo{author}{Bryant, P.}, \bibinfo{author}{Pozzati, G.} \& \bibinfo{author}{Elofsson, A.}
\newblock \bibinfo{journal}{\bibinfo{title}{Improved prediction of protein-protein interactions using alphafold2}}.
\newblock {\emph{\JournalTitle{Nature communications}}} \textbf{\bibinfo{volume}{13}}, \bibinfo{pages}{1265} (\bibinfo{year}{2022}).

\bibitem{baek2021accurate}
\bibinfo{author}{Baek, M.} \emph{et~al.}
\newblock \bibinfo{journal}{\bibinfo{title}{Accurate prediction of protein structures and interactions using a three-track neural network}}.
\newblock {\emph{\JournalTitle{Science}}} \textbf{\bibinfo{volume}{373}}, \bibinfo{pages}{871--876} (\bibinfo{year}{2021}).

\bibitem{dobson2003protein_fold}
\bibinfo{author}{Dobson, C.~M.}
\newblock \bibinfo{journal}{\bibinfo{title}{Protein folding and misfolding}}.
\newblock {\emph{\JournalTitle{Nature}}} \textbf{\bibinfo{volume}{426}}, \bibinfo{pages}{884--890} (\bibinfo{year}{2003}).

\bibitem{alex_krizhevsky2012imagenet}
\bibinfo{author}{Krizhevsky, A.}, \bibinfo{author}{Sutskever, I.} \& \bibinfo{author}{Hinton, G.~E.}
\newblock \bibinfo{journal}{\bibinfo{title}{Imagenet classification with deep convolutional neural networks}}.
\newblock {\emph{\JournalTitle{Advances in neural information processing systems}}} \textbf{\bibinfo{volume}{25}} (\bibinfo{year}{2012}).

\bibitem{self_driving_bojarski2016end}
\bibinfo{author}{Bojarski, M.} \emph{et~al.}
\newblock \bibinfo{journal}{\bibinfo{title}{End to end learning for self-driving cars}}.
\newblock {\emph{\JournalTitle{arXiv preprint arXiv:1604.07316}}}  (\bibinfo{year}{2016}).

\bibitem{amodei2016deep_speech}
\bibinfo{author}{Amodei, D.} \emph{et~al.}
\newblock \bibinfo{title}{Deep speech 2: End-to-end speech recognition in english and mandarin}.
\newblock In \emph{\bibinfo{booktitle}{International conference on machine learning}}, \bibinfo{pages}{173--182} (\bibinfo{organization}{PMLR}, \bibinfo{year}{2016}).

\bibitem{liu;aca19}
\bibinfo{author}{Liu, C.-H.}, \bibinfo{author}{Tao, Y.}, \bibinfo{author}{Hsu, D.}, \bibinfo{author}{Du, Q.} \& \bibinfo{author}{Billinge, S. J.~L.}
\newblock \bibinfo{journal}{\bibinfo{title}{Using a machine learning approach to determine the space group of a structure from the atomic pair distribution function}}.
\newblock {\emph{\JournalTitle{Acta Cryst A}}} \textbf{\bibinfo{volume}{75}}, \bibinfo{pages}{633--643}, \doiprefix\url{10.1107/S2053273319005606} (\bibinfo{year}{2019}).

\bibitem{xrd_class_oviedo2019fast}
\bibinfo{author}{Oviedo, F.} \emph{et~al.}
\newblock \bibinfo{journal}{\bibinfo{title}{Fast and interpretable classification of small x-ray diffraction datasets using data augmentation and deep neural networks}}.
\newblock {\emph{\JournalTitle{npj Computational Materials}}} \textbf{\bibinfo{volume}{5}}, \bibinfo{pages}{60} (\bibinfo{year}{2019}).

\bibitem{xrd_class_ml_suzuki2020symmetry}
\bibinfo{author}{Suzuki, Y.} \emph{et~al.}
\newblock \bibinfo{journal}{\bibinfo{title}{Symmetry prediction and knowledge discovery from x-ray diffraction patterns using an interpretable machine learning approach}}.
\newblock {\emph{\JournalTitle{Scientific reports}}} \textbf{\bibinfo{volume}{10}}, \bibinfo{pages}{21790} (\bibinfo{year}{2020}).

\bibitem{xrd_park2017classification}
\bibinfo{author}{Park, W.~B.} \emph{et~al.}
\newblock \bibinfo{journal}{\bibinfo{title}{Classification of crystal structure using a convolutional neural network}}.
\newblock {\emph{\JournalTitle{IUCrJ}}} \textbf{\bibinfo{volume}{4}}, \bibinfo{pages}{486--494} (\bibinfo{year}{2017}).

\bibitem{xrd_lee2020deep}
\bibinfo{author}{Lee, J.-W.}, \bibinfo{author}{Park, W.~B.}, \bibinfo{author}{Lee, J.~H.}, \bibinfo{author}{Singh, S.~P.} \& \bibinfo{author}{Sohn, K.-S.}
\newblock \bibinfo{journal}{\bibinfo{title}{A deep-learning technique for phase identification in multiphase inorganic compounds using synthetic xrd powder patterns}}.
\newblock {\emph{\JournalTitle{Nature communications}}} \textbf{\bibinfo{volume}{11}}, \bibinfo{pages}{86} (\bibinfo{year}{2020}).

\bibitem{xrd_aguiar2019decoding}
\bibinfo{author}{Aguiar, J.}, \bibinfo{author}{Gong, M.~L.}, \bibinfo{author}{Unocic, R.}, \bibinfo{author}{Tasdizen, T.} \& \bibinfo{author}{Miller, B.}
\newblock \bibinfo{journal}{\bibinfo{title}{Decoding crystallography from high-resolution electron imaging and diffraction datasets with deep learning}}.
\newblock {\emph{\JournalTitle{Science Advances}}} \textbf{\bibinfo{volume}{5}}, \bibinfo{pages}{eaaw1949} (\bibinfo{year}{2019}).

\bibitem{class_ziletti2018insightful}
\bibinfo{author}{Ziletti, A.}, \bibinfo{author}{Kumar, D.}, \bibinfo{author}{Scheffler, M.} \& \bibinfo{author}{Ghiringhelli, L.~M.}
\newblock \bibinfo{journal}{\bibinfo{title}{Insightful classification of crystal structures using deep learning}}.
\newblock {\emph{\JournalTitle{Nature communications}}} \textbf{\bibinfo{volume}{9}}, \bibinfo{pages}{2775} (\bibinfo{year}{2018}).

\bibitem{xrd_tiong2020identification}
\bibinfo{author}{Tiong, L. C.~O.}, \bibinfo{author}{Kim, J.}, \bibinfo{author}{Han, S.~S.} \& \bibinfo{author}{Kim, D.}
\newblock \bibinfo{journal}{\bibinfo{title}{Identification of crystal symmetry from noisy diffraction patterns by a shape analysis and deep learning}}.
\newblock {\emph{\JournalTitle{npj Computational Materials}}} \textbf{\bibinfo{volume}{6}}, \bibinfo{pages}{196} (\bibinfo{year}{2020}).

\bibitem{garcia2019learning}
\bibinfo{author}{Garcia-Cardona, C.} \emph{et~al.}
\newblock \bibinfo{title}{Learning to predict material structure from neutron scattering data}.
\newblock In \emph{\bibinfo{booktitle}{2019 IEEE International Conference on Big Data (Big Data)}}, \bibinfo{pages}{4490--4497} (\bibinfo{organization}{IEEE}, \bibinfo{year}{2019}).

\bibitem{merker2022machine}
\bibinfo{author}{Merker, H.~A.} \emph{et~al.}
\newblock \bibinfo{journal}{\bibinfo{title}{Machine learning magnetism classifiers from atomic coordinates}}.
\newblock {\emph{\JournalTitle{Iscience}}} \textbf{\bibinfo{volume}{25}} (\bibinfo{year}{2022}).

\bibitem{deepmind_generative_merchant2023scaling}
\bibinfo{author}{Merchant, A.} \emph{et~al.}
\newblock \bibinfo{journal}{\bibinfo{title}{Scaling deep learning for materials discovery}}.
\newblock {\emph{\JournalTitle{Nature}}} \bibinfo{pages}{1--6} (\bibinfo{year}{2023}).

\bibitem{materials_diffusion_yang2023scalable}
\bibinfo{author}{Yang, M.} \emph{et~al.}
\newblock \bibinfo{journal}{\bibinfo{title}{Scalable diffusion for materials discovery}}.
\newblock {\emph{\JournalTitle{arXiv e-prints}}}  (\bibinfo{year}{2023}).

\bibitem{gflownet_hernandez2023crystal}
\bibinfo{author}{Hern{\'a}ndez-Garc{\'\i}a, A.} \emph{et~al.}
\newblock \bibinfo{title}{Crystal-gflownet: sampling materials with desirable properties and constraints}.
\newblock In \emph{\bibinfo{booktitle}{AI for Accelerated Materials Design-NeurIPS 2023 Workshop}} (\bibinfo{year}{2023}).

\bibitem{pan2023deep}
\bibinfo{author}{Pan, T.}, \bibinfo{author}{Jin, S.}, \bibinfo{author}{Miller, M.~D.}, \bibinfo{author}{Kyrillidis, A.} \& \bibinfo{author}{Phillips, G.~N.}
\newblock \bibinfo{journal}{\bibinfo{title}{A deep learning solution for crystallographic structure determination}}.
\newblock {\emph{\JournalTitle{IUCrJ}}} \textbf{\bibinfo{volume}{10}}, \bibinfo{pages}{487--496} (\bibinfo{year}{2023}).

\bibitem{dun2023crysformer}
\bibinfo{author}{Dun, C.} \emph{et~al.}
\newblock \bibinfo{journal}{\bibinfo{title}{Crysformer: Protein structure prediction via 3d patterson maps and partial structure attention}}.
\newblock {\emph{\JournalTitle{arXiv preprint arXiv:2310.03899}}}  (\bibinfo{year}{2023}).

\bibitem{alphafold_crystallography_barbarin2022x}
\bibinfo{author}{Barbarin-Bocahu, I.} \& \bibinfo{author}{Graille, M.}
\newblock \bibinfo{journal}{\bibinfo{title}{The x-ray crystallography phase problem solved thanks to alphafold and rosettafold models: a case-study report}}.
\newblock {\emph{\JournalTitle{Acta Crystallographica Section D: Structural Biology}}} \textbf{\bibinfo{volume}{78}}, \bibinfo{pages}{517--531} (\bibinfo{year}{2022}).

\bibitem{kjaerDeepStrucStructureSolution2023a}
\bibinfo{author}{Kj{\ae}r, E. T.~S.} \emph{et~al.}
\newblock \bibinfo{journal}{\bibinfo{title}{{{DeepStruc}}: Towards structure solution from pair distribution function data using deep generative models}}.
\newblock {\emph{\JournalTitle{Digital Discovery}}} \textbf{\bibinfo{volume}{2}}, \bibinfo{pages}{69--80}, \doiprefix\url{10.1039/D2DD00086E} (\bibinfo{year}{2023}).

\bibitem{kingma2013auto_vae}
\bibinfo{author}{Kingma, D.~P.} \& \bibinfo{author}{Welling, M.}
\newblock \bibinfo{journal}{\bibinfo{title}{Auto-encoding variational bayes}}.
\newblock {\emph{\JournalTitle{arXiv preprint arXiv:1312.6114}}}  (\bibinfo{year}{2013}).

\bibitem{yu2021pixelnerf}
\bibinfo{author}{Yu, A.}, \bibinfo{author}{Ye, V.}, \bibinfo{author}{Tancik, M.} \& \bibinfo{author}{Kanazawa, A.}
\newblock \bibinfo{title}{pixelnerf: Neural radiance fields from one or few images}.
\newblock In \emph{\bibinfo{booktitle}{Proceedings of the IEEE/CVF Conference on Computer Vision and Pattern Recognition}}, \bibinfo{pages}{4578--4587} (\bibinfo{year}{2021}).

\bibitem{mildenhall2021nerf}
\bibinfo{author}{Mildenhall, B.} \emph{et~al.}
\newblock \bibinfo{journal}{\bibinfo{title}{Nerf: Representing scenes as neural radiance fields for view synthesis}}.
\newblock {\emph{\JournalTitle{Communications of the ACM}}} \textbf{\bibinfo{volume}{65}}, \bibinfo{pages}{99--106} (\bibinfo{year}{2021}).

\bibitem{tancik2020fourier}
\bibinfo{author}{Tancik, M.} \emph{et~al.}
\newblock \bibinfo{journal}{\bibinfo{title}{Fourier features let networks learn high frequency functions in low dimensional domains}}.
\newblock {\emph{\JournalTitle{Advances in Neural Information Processing Systems}}} \textbf{\bibinfo{volume}{33}}, \bibinfo{pages}{7537--7547} (\bibinfo{year}{2020}).

\bibitem{sitzmann2019scene}
\bibinfo{author}{Sitzmann, V.}, \bibinfo{author}{Zollh{\"o}fer, M.} \& \bibinfo{author}{Wetzstein, G.}
\newblock \bibinfo{journal}{\bibinfo{title}{Scene representation networks: Continuous 3d-structure-aware neural scene representations}}.
\newblock {\emph{\JournalTitle{Advances in Neural Information Processing Systems}}} \textbf{\bibinfo{volume}{32}} (\bibinfo{year}{2019}).

\bibitem{park2019deepsdf}
\bibinfo{author}{Park, J.~J.}, \bibinfo{author}{Florence, P.}, \bibinfo{author}{Straub, J.}, \bibinfo{author}{Newcombe, R.} \& \bibinfo{author}{Lovegrove, S.}
\newblock \bibinfo{title}{Deepsdf: Learning continuous signed distance functions for shape representation}.
\newblock In \emph{\bibinfo{booktitle}{Proceedings of the IEEE/CVF conference on computer vision and pattern recognition}}, \bibinfo{pages}{165--174} (\bibinfo{year}{2019}).

\bibitem{hoffmannDataDrivenApproachEncoding2019}
\bibinfo{author}{Hoffmann, J.} \emph{et~al.}
\newblock \bibinfo{title}{Data-{{Driven Approach}} to {{Encoding}} and {{Decoding}} 3-{{D Crystal Structures}}} (\bibinfo{year}{2019}).
\newblock \eprint{1909.00949}.

\bibitem{higgins2016beta_vae}
\bibinfo{author}{Higgins, I.} \emph{et~al.}
\newblock \bibinfo{title}{beta-vae: Learning basic visual concepts with a constrained variational framework}.
\newblock In \emph{\bibinfo{booktitle}{International conference on learning representations}} (\bibinfo{year}{2016}).

\bibitem{materials_project_jain2013commentary}
\bibinfo{author}{Jain, A.} \emph{et~al.}
\newblock \bibinfo{journal}{\bibinfo{title}{Commentary: The materials project: A materials genome approach to accelerating materials innovation}}.
\newblock {\emph{\JournalTitle{APL materials}}} \textbf{\bibinfo{volume}{1}} (\bibinfo{year}{2013}).

\bibitem{ssim_wang2004image}
\bibinfo{author}{Wang, Z.}, \bibinfo{author}{Bovik, A.~C.}, \bibinfo{author}{Sheikh, H.~R.} \& \bibinfo{author}{Simoncelli, E.~P.}
\newblock \bibinfo{journal}{\bibinfo{title}{Image quality assessment: from error visibility to structural similarity}}.
\newblock {\emph{\JournalTitle{IEEE transactions on image processing}}} \textbf{\bibinfo{volume}{13}}, \bibinfo{pages}{600--612} (\bibinfo{year}{2004}).

\bibitem{psnr_hore2010image}
\bibinfo{author}{Hore, A.} \& \bibinfo{author}{Ziou, D.}
\newblock \bibinfo{title}{Image quality metrics: Psnr vs. ssim}.
\newblock In \emph{\bibinfo{booktitle}{2010 20th international conference on pattern recognition}}, \bibinfo{pages}{2366--2369} (\bibinfo{organization}{IEEE}, \bibinfo{year}{2010}).

\bibitem{icsd_belsky2002new}
\bibinfo{author}{Belsky, A.}, \bibinfo{author}{Hellenbrandt, M.}, \bibinfo{author}{Karen, V.~L.} \& \bibinfo{author}{Luksch, P.}
\newblock \bibinfo{journal}{\bibinfo{title}{New developments in the inorganic crystal structure database (icsd): accessibility in support of materials research and design}}.
\newblock {\emph{\JournalTitle{Acta Crystallographica Section B: Structural Science}}} \textbf{\bibinfo{volume}{58}}, \bibinfo{pages}{364--369} (\bibinfo{year}{2002}).

\bibitem{vasp_hafner2008ab}
\bibinfo{author}{Hafner, J.}
\newblock \bibinfo{journal}{\bibinfo{title}{Ab-initio simulations of materials using vasp: Density-functional theory and beyond}}.
\newblock {\emph{\JournalTitle{Journal of computational chemistry}}} \textbf{\bibinfo{volume}{29}}, \bibinfo{pages}{2044--2078} (\bibinfo{year}{2008}).

\bibitem{materials_project_extra_info}
\bibinfo{title}{The materials project workshop}.
\newblock \bibinfo{howpublished}{\url{https://workshop.materialsproject.org/lessons/01_website_walkthrough/website_walkthrough/}}.
\newblock \bibinfo{note}{Retrieved 08/25/2023}.

\bibitem{diffraction_calculation_de2012structure}
\bibinfo{author}{De~Graef, M.} \& \bibinfo{author}{McHenry, M.~E.}
\newblock \emph{\bibinfo{title}{Structure of materials: an introduction to crystallography, diffraction and symmetry}} (\bibinfo{publisher}{Cambridge University Press}, \bibinfo{year}{2012}).

\bibitem{materials_project_diffraction_calculation}
\bibinfo{title}{Diffraction patterns: How diffraction patterns are calculated on the materials project (mp) website.}
\newblock \bibinfo{howpublished}{\url{https://docs.materialsproject.org/methodology/materials-methodology/diffraction-patterns}}.
\newblock \bibinfo{note}{Retrieved 08/25/2023}.

\bibitem{ioffe2015batchnorm}
\bibinfo{author}{Ioffe, S.} \& \bibinfo{author}{Szegedy, C.}
\newblock \bibinfo{title}{Batch normalization: Accelerating deep network training by reducing internal covariate shift}.
\newblock In \emph{\bibinfo{booktitle}{International conference on machine learning}}, \bibinfo{pages}{448--456} (\bibinfo{organization}{pmlr}, \bibinfo{year}{2015}).

\bibitem{materials_project_charge_density_calculation}
\bibinfo{title}{Charge density: Obtaining the charge density shown on the materials project (mp) website.}
\newblock \bibinfo{howpublished}{\url{https://docs.materialsproject.org/methodology/materials-methodology/charge-density}}.
\newblock \bibinfo{note}{Retrieved 08/25/2023}.

\bibitem{vasp_chgcar}
\bibinfo{title}{Chgcar}.
\newblock \bibinfo{howpublished}{\url{https://www.vasp.at/wiki/index.php/CHGCAR}}.
\newblock \bibinfo{note}{Retrieved 08/25/2023}.

\bibitem{pyrho_shen2022representation}
\bibinfo{author}{Shen, J.-X.} \emph{et~al.}
\newblock \bibinfo{journal}{\bibinfo{title}{A representation-independent electronic charge density database for crystalline materials}}.
\newblock {\emph{\JournalTitle{Scientific Data}}} \textbf{\bibinfo{volume}{9}}, \bibinfo{pages}{661} (\bibinfo{year}{2022}).

\bibitem{pyrho}
\bibinfo{title}{mp-pyrho}.
\newblock \bibinfo{howpublished}{\url{https://github.com/materialsproject/pyrho}}.
\newblock \bibinfo{note}{Retrieved 08/25/2023}.

\bibitem{lattice_prediction_from_xrd_chitturi2021automated}
\bibinfo{author}{Chitturi, S.~R.} \emph{et~al.}
\newblock \bibinfo{journal}{\bibinfo{title}{Automated prediction of lattice parameters from x-ray powder diffraction patterns}}.
\newblock {\emph{\JournalTitle{Journal of Applied Crystallography}}} \textbf{\bibinfo{volume}{54}}, \bibinfo{pages}{1799--1810} (\bibinfo{year}{2021}).

\bibitem{guccioneExtractionCrystalCell2023}
\bibinfo{author}{Guccione, P.}, \bibinfo{author}{Diacono, D.}, \bibinfo{author}{Toso, S.} \& \bibinfo{author}{Caliandro, R.}
\newblock \bibinfo{journal}{\bibinfo{title}{Towards the extraction of the crystal cell parameters from pair distribution function profiles}}.
\newblock {\emph{\JournalTitle{IUCrJ}}} \textbf{\bibinfo{volume}{10}}, \bibinfo{pages}{610--623}, \doiprefix\url{10.1107/S2052252523006887} (\bibinfo{year}{2023}).

\bibitem{densenet_huang2017densely}
\bibinfo{author}{Huang, G.}, \bibinfo{author}{Liu, Z.}, \bibinfo{author}{Van Der~Maaten, L.} \& \bibinfo{author}{Weinberger, K.~Q.}
\newblock \bibinfo{title}{Densely connected convolutional networks}.
\newblock In \emph{\bibinfo{booktitle}{Proceedings of the IEEE conference on computer vision and pattern recognition}}, \bibinfo{pages}{4700--4708} (\bibinfo{year}{2017}).

\bibitem{ba2016layernorm}
\bibinfo{author}{Ba, J.~L.}, \bibinfo{author}{Kiros, J.~R.} \& \bibinfo{author}{Hinton, G.~E.}
\newblock \bibinfo{journal}{\bibinfo{title}{Layer normalization}}.
\newblock {\emph{\JournalTitle{arXiv preprint arXiv:1607.06450}}}  (\bibinfo{year}{2016}).

\bibitem{jacot2018ntk}
\bibinfo{author}{Jacot, A.}, \bibinfo{author}{Gabriel, F.} \& \bibinfo{author}{Hongler, C.}
\newblock \bibinfo{journal}{\bibinfo{title}{Neural tangent kernel: Convergence and generalization in neural networks}}.
\newblock {\emph{\JournalTitle{Advances in neural information processing systems}}} \textbf{\bibinfo{volume}{31}} (\bibinfo{year}{2018}).

\bibitem{perez2018film}
\bibinfo{author}{Perez, E.}, \bibinfo{author}{Strub, F.}, \bibinfo{author}{De~Vries, H.}, \bibinfo{author}{Dumoulin, V.} \& \bibinfo{author}{Courville, A.}
\newblock \bibinfo{title}{Film: Visual reasoning with a general conditioning layer}.
\newblock In \emph{\bibinfo{booktitle}{Proceedings of the AAAI conference on artificial intelligence}}, vol.~\bibinfo{volume}{32} (\bibinfo{year}{2018}).

\bibitem{kl_div_kullback1951information}
\bibinfo{author}{Kullback, S.} \& \bibinfo{author}{Leibler, R.~A.}
\newblock \bibinfo{journal}{\bibinfo{title}{On information and sufficiency}}.
\newblock {\emph{\JournalTitle{The annals of mathematical statistics}}} \textbf{\bibinfo{volume}{22}}, \bibinfo{pages}{79--86} (\bibinfo{year}{1951}).

\bibitem{kingma2014adam}
\bibinfo{author}{Kingma, D.~P.} \& \bibinfo{author}{Ba, J.}
\newblock \bibinfo{journal}{\bibinfo{title}{Adam: A method for stochastic optimization}}.
\newblock {\emph{\JournalTitle{arXiv preprint arXiv:1412.6980}}}  (\bibinfo{year}{2014}).

\bibitem{loshchilov2016sgdr_cosine_warm_restart}
\bibinfo{author}{Loshchilov, I.} \& \bibinfo{author}{Hutter, F.}
\newblock \bibinfo{journal}{\bibinfo{title}{Sgdr: Stochastic gradient descent with warm restarts}}.
\newblock {\emph{\JournalTitle{arXiv preprint arXiv:1608.03983}}}  (\bibinfo{year}{2016}).

\bibitem{van2014scikit_image}
\bibinfo{author}{Van~der Walt, S.} \emph{et~al.}
\newblock \bibinfo{journal}{\bibinfo{title}{scikit-image: image processing in python}}.
\newblock {\emph{\JournalTitle{PeerJ}}} \textbf{\bibinfo{volume}{2}}, \bibinfo{pages}{e453} (\bibinfo{year}{2014}).

\end{thebibliography}

\section*{Acknowledgements}

Work in the Lipson group was supported by U.S. National Science Foundation under AI Institute for Dynamical Systems grant 2112085. Work in the Billinge group was supported by the U.S. Department of Energy, Office of Science, Office of Basic Energy Sciences (DOE-BES) under contract No. DE-SC0024141. 

\section*{Author contributions statement}

HL and SJLB proposed the research. GG, BC, and HL designed the system architecture. GG, SJLB, and HL designed the experiment systems. GG wrote the code. JG, AHY, and AR ensured the quality of the code. BC and LL conducted initial explorations. JG, LL, and AR provided helpful insights.  SJLB, HL, BC, JG, LL, and GG wrote the paper.

% \section*{Additional information}
%\newpage

%\input{supplement}

\end{document}